\documentclass[twocolumn, trackchanges]{aastex62}

\usepackage{multirow}
\usepackage{asmath}
\usepackage{graphicx}
\usepackage{bm}
\usepackage{catchfile}

\newcommand{\myemail}{cornachi@usna.edu}
\newcommand{\Romannumeral}[1]{\uppercase\expandafter{\romannumeral #1\relax}}

\shorttitle{WFI2026 Microlensing Size}
\shortauthors{Cornachione et al.}
\accepted{Nov. 6, 2019}

\begin{document}

\title{A Microlensing Accretion Disk Size Measurement in the Lensed Quasar WFI 2026--4536}

\author[0000-0003-1012-4771]{Matthew A. Cornachione}
\affiliation{Department of Physics, United States Naval Academy, 572C Holloway Rd., Annapolis, MD 21402, USA}
\email{\myemail}

\author[0000-0003-2460-9999]{Christopher W. Morgan}
\affiliation{Department of Physics, United States Naval Academy, 572C Holloway Rd., Annapolis, MD 21402, USA}

\author{Martin Millon}
\affiliation{Institute of Physics, Laboratoire d' Astrophysique, \'{E}cole Polytechnique F\'ed\'erale de Lausanne (EPFL), Observatoire de Sauverny, 1290 Versoix, Switzerland}

\author[0000-0002-2816-5398]{Misty C. Bentz}
\affiliation{Department of Physics and Astronomy,
		 Georgia State University,
		 Atlanta, GA 30303, USA;
		 bentz@astro.gsu.edu}

\author{Frederic Courbin}
\affiliation{Institute of Physics, Laboratoire d' Astrophysique, \'{E}cole Polytechnique F\'ed\'erale de Lausanne (EPFL), Observatoire de Sauverny, 1290 Versoix, Switzerland}

\author[0000-0003-1471-3952]{Vivien Bonvin}
\affiliation{Institute of Physics, Laboratoire d' Astrophysique, \'{E}cole Polytechnique F\'ed\'erale de Lausanne (EPFL), Observatoire de Sauverny, 1290 Versoix, Switzerland}

\author[0000-0002-7061-6519]{Emilio E. Falco}
\affiliation{Harvard-Smithsonian Center for Astrophysics, 60 Garden St., Cambridge, MA 02138, USA}

\begin{abstract}
We use thirteen seasons of R-band photometry from the 1.2m Leonard Euler Swiss Telescope at La Silla to examine microlensing variability in the quadruply-imaged lensed quasar WFI 2026--4536.  The lightcurves exhibit ${\sim}\,0.2\,\text{mag}$ of uncorrelated variability across all epochs and a prominent single feature of ${\sim}\,0.1\,\text{mag}$ within a single season. We analyze this variability to constrain the size of the quasar's accretion disk.  Adopting a nominal inclination of 60\degr, we find an accretion disk scale radius of $\log(r_s/\text{cm}) = 15.74^{+0.34}_{-0.29}$ at a rest-frame wavelength of $2043\,\text{\AA}$, and we estimate a black hole mass of $\log(M_{\text{BH}}/M_{\odot}) = 9.18^{+0.39}_{-0.34}$, based on the \ion{C}{4} line in VLT spectra.  This size measurement is fully consistent with the Quasar Accretion Disk Size -- Black Hole Mass relation, providing another system in which the accretion disk is larger than predicted by thin disk theory. 
\end{abstract}

\section{Introduction}
\label{sec:intro}

Quasars, a class of active galactic nuclei (AGN) \citep[e.g.,][]{anto1993a, urry1995a}, have been an area of intense study for decades.  However, their small physical sizes subtend angles that are much smaller than the resolution limit of any existing telescope.  Hence we have been forced to infer the physics powering these luminous sources by studying intrinsic flux variability \citep[e.g.,][]{vand2004a, serg2005a, cack2007a, kell2009a, macl2010a}, modeling spectral profiles \citep{sun1989a, bonn2007a, gask2008a, hall2018a}, or using reverberation mapping \citep[e.g.,][]{pete2004a, bent2010a, edel2015a, cack2018a}.  While these methods have provided insights into quasar structure and central black hole masses, the accretion disk continuum size and temperature profile remain open research areas.

Microlensing, first observed by \citet{chan1979a}, has offered an opportunity to better measure the size of quasars.  Strongly lensed quasar images are magnified by a complex field of stellar-mass objects in the lens galaxy.  As the quasar moves relative to our line of sight, the magnification changes, generating significant uncorrelated variability between images on timescales of months to years.   If the time delays between images are known, it is possible to distinguish the correlated instrinsic quasar variability from the uncorrelated microlensing variability. \citet{koch2004a} developed a Bayesian Monte Carlo technique to measure the sizes of quasars from multiple-epoch lightcurves. 
With this technique, we have made measurements of accretion disk scale sizes in 15 quasars \citep{koch2006a, morg2006a, poin2007a, morg2008a, morg2010a, morg2012a, hain2012a, hain2013a, mosq2013a, blac2014a, macl2015a, morg2018a}.  This method requires cosmological modeling of the effective transverse velocity, but is insensitive to uncertainty in the median mass of stars in the lens galaxy.   A machine learning analysis technique, developed by \citet{vern2019a}, may allow for more rapid analysis of larger sets of quasar lightcurves.

Alternatively, microlensing sizes can be inferred from chromatic variation between lensed images.  In this method a quasar is imaged at a single-epoch across multiple filter bands.  This approach has generated complementary measurements of quasar accretion disk sizes \citep{pool2007a, bate2008a, blac2011a, medi2011a, mosq2011b, pool2012a, jime2012a, sche2014a, mott2017a, bate2018a}.  This method uses dramatically less observing time, but requires careful treatment of broad emission line contamination and flux offsets due to dust or millilensing.  Furthermore, all reported sizes are subject to an assumed prior on the unknown median mass of stars in the lens galaxy.  Nevertheless, when combined with constraints from multi-epoch studies, the single-epoch method generally gives similar accretion disk size measurements.

Because of the rarity of lensed quasar discoveries and the onerous observing requirements of multi-epoch studies, only fourteen multi-epoch size measurements have been reported to date \citep[and references therein]{morg2018a}.  Here we increase that number to 15 with the addition of the quadruply lensed WFI J2026--4536 (hereafter WFI2026)\footnote{Based on observations made with the ESO-VLT Unit Telescope 2 Kueyen (Cerro Paranal, Chile; Programs 074.A-0563 and 075.A-0377, PI: G. Meylan} \citep{morg2004a}.  The quasar source is at redshift $z_s = 2.23$, but the lens redshift was not measurable in archival spectra from the Very Large Telescope (VLT).  The lens galaxy is faint and stacked galaxy spectra from fourteen exposures show no distinct spectral features.  Although we were unable to estimate the lens redshift, we succeeded in using these archival VLT spectra to measure the black hole mass.

 Two different investigations have already used the single-epoch technique to estimate the accretion disk size in this system. 
\citet{blac2011a} observed WFI2026 in the infrared using the Persson's Auxiliary Nasmyth Infrared Camera (PANIC) on the Baade telescope and in the optical using the Raymond and Beverly Sackler Magellan Instant Camera (MagIC) on both the Clay and Baade telescopes at Las Campanas Observatory.  They estimated an accretion disk half-light radius of $\log (r_{1/2}/\text{cm}) = 16.46 \pm 0.32$ at $2043\text{\AA}$, under a log-prior on $r_{1/2}$.  A recent analysis by \citet{bate2018a} using IR and UVIS channels on the Wide Field Camera 3 (WFC3) on the Hubble Space Telescope (\textsl{HST}) found evidence for a smaller size, $\log (r_{1/2}/\text{cm}) < 16$ (re-scaled to $2043\text{\AA}$ from an observed $1026\text{\AA}$).  Both of these estimates assume a $0.3 M_{\odot}$ median stellar mass in the lens galaxy.

Analysis of our 13 season lightcurve complements these previous studies with a multi-epoch constraint on the scale radius, $r_s$, and addresses the mild tension between previous results.
Note that these studies reported the measured size as a half light radius, $r_{1/2}$, while we cast our results as a thin disk scale radius, $r_s = r_{1/2}/2.44$, to facilitate comparison with theoretical disk models.  In any case, \citet{mort2005a} showed that projected area, not shape, dominates microlensing variablility, so these radii are directly proportional with $r_s = r_{1/2}/a$.  The scaling factor $a$ depends on the assumed disk geometry with $a=2.44$ for a thin disk and $a=1.18$ for a Gaussian disk.

This paper is organized as follows.  In section~\ref{sec:data}, we present our monitoring data, photometric technique, and the reduced light curves.  In section~\ref{sec:lens_model} we present our strong lens modeling and our determination of time delays for the system.  We discuss our microlensing model in section~\ref{sec:ml_model}, and we present our measurements for the WFI2026 disk size and black hole mass in section~\ref{sec:results}. 
In section~\ref{sec:discussion} we compare our measurements to those of \citet{blac2011a} and \citet{bate2018a} and we conclude with a discussion of the accretion disk size and black hole mass in the context of previous multi-epoch studies.

\section{Data}
\label{sec:data}

\begin{figure*}[t]
\gridline{\fig{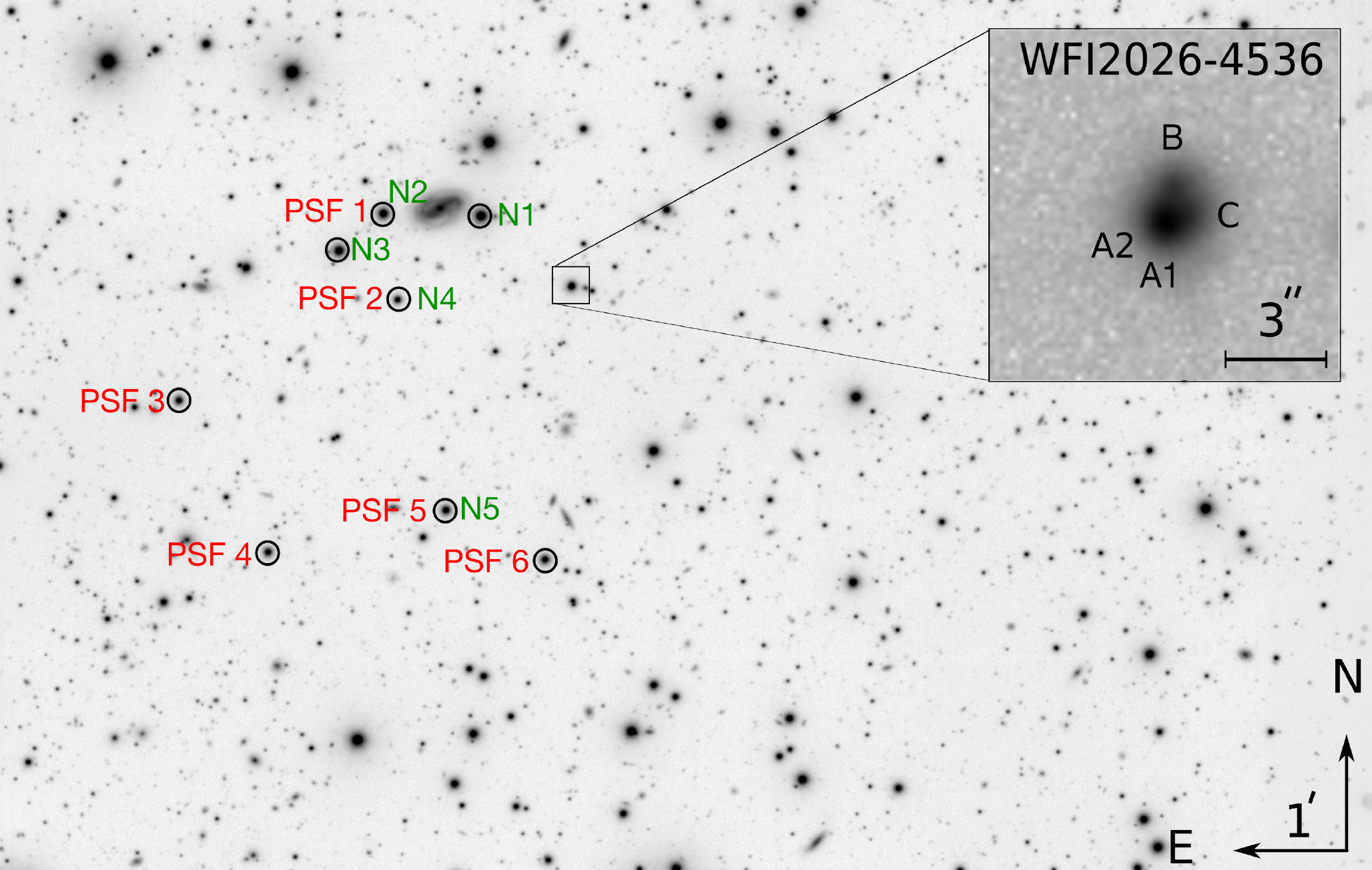}{0.8 \textwidth}{}}
\caption{Deep field stack of reduced images of WFI2026 from the ECAM detector.  Stars used for fitting the point spread function (PSF) are labeled in red and stars used for flux normalization (N) are labeled in green.  The relative positions of the lensed quasar images are indicated in the expanded box, showing a single subexposure in excellent seeing.}
\label{fig:raw_ccds}
\end{figure*}

The observational campaign was conducted within the scope of the COSmological MOnitoring of GRAvItational Lenses COSMOGRAIL collaboration \citep[e.g.][]{cour2005a, bonv2018a}  We used images of WFI2026 obtained with the Swiss 1.2m Leonhard Euler telescope (hereafter Euler) located at La Silla Observatory in Chile between April, 2004 and November, 2016.  Prior to 2010 we collected data using the C2 chip, a $2048\times2048$ detector with a pixel scale of $0\farcs344$.  We took more recent exposures on the EulerCAM (ECAM) detector, with a smaller pixel scale of $0\farcs2149$ and dimensions of $3496\times3512$ pixels.  Across all epochs we used the Rouge Gen\'{e}ve (RG) filter, a modified R-band filter with an effective wavelength of $6600\,\text{\AA}$.  At each of the 548 epochs, we obtained five $360\,\text{s}$ subexposures.  Our observations with Euler spanned 13 seasons with a typical observation cadence of once every six days for C2 and once every four days for ECAM with inter-season gaps of ${\sim}\,100$ days.  In Figure~\ref{fig:raw_ccds} we show a stacked ECAM exposure of WFI2026 indicating the reference stars used for point spread function (PSF) calibration and flux normalization (N).

\begin{figure*}[t]
\gridline{\fig{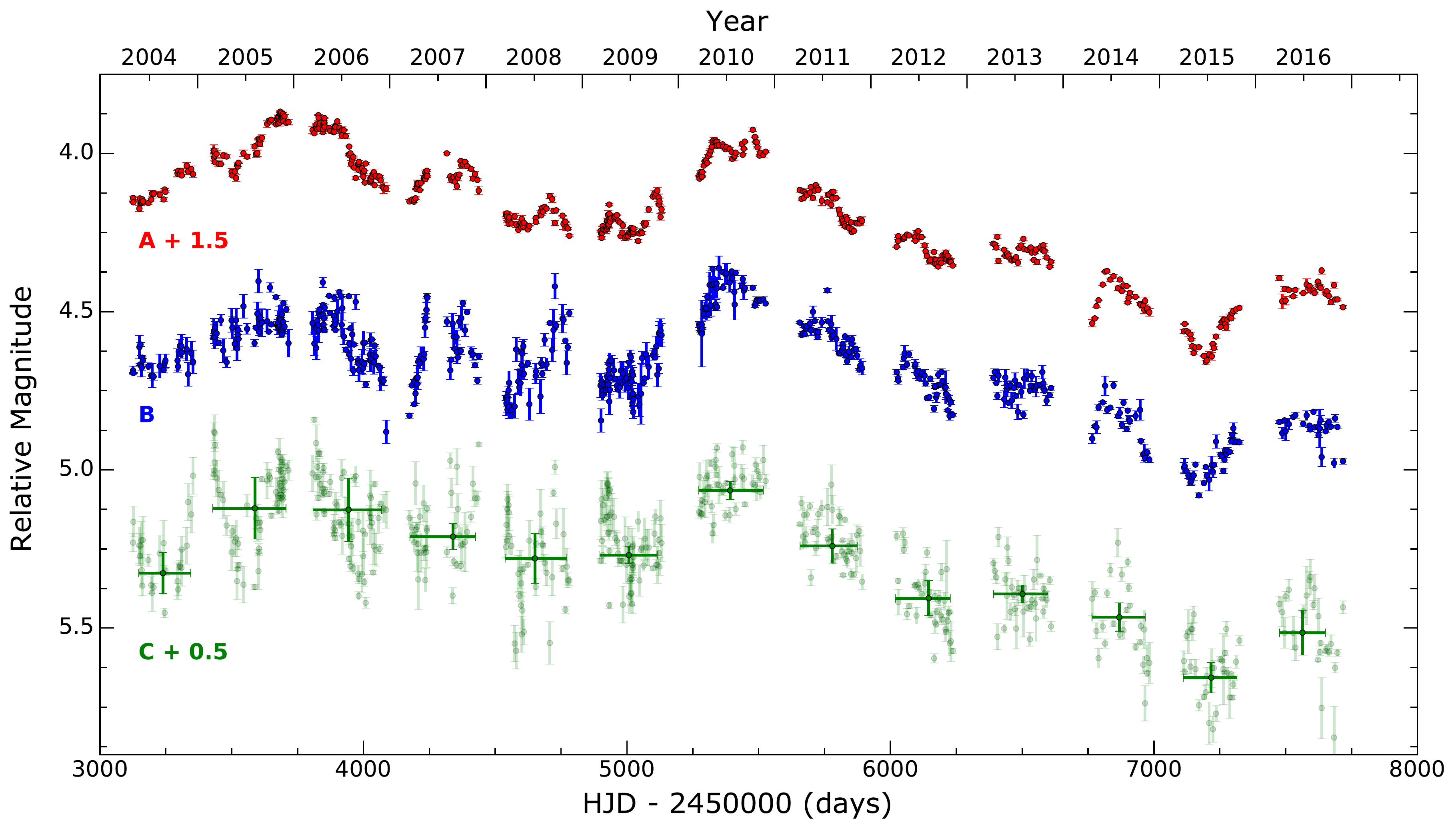}{1.0 \textwidth}{}}
\caption{Reduced lightcurve for WFI2026 obtained with Euler.  The curves are plotted in relative magnitudes with an arbitrary offset.  Image A (A1+A2) is red, image B is blue, and image C is green.  The season averages for image C are overlaid atop the reduced lightcurve.}
\label{fig:lightcurves}
\end{figure*}

\begin{deluxetable*}{c|c|c|c|c|c|c}
\tablecaption{Astrometric measurements for WFI2026 based on \textsl{HST} CASTLES imaging, F160W band.  We used image B as the position reference.  The lens galaxy is indicated by G.  The effective radius, $r_e$, ellipticity, $e$, and position angle, $\theta_e$, are indicated for the galaxy using a de Vaucouleurs profile.\label{tab:hst_astrom}}
\tablehead{
\colhead{Component} & \colhead{$\Delta$RA (\arcsec)} & \colhead{$\Delta$Dec (\arcsec)} & \colhead{$r_e$ (\arcsec)} & \colhead{$e$} & \colhead{$\theta_e$ (\degr)} & \colhead{F160W mag}
}
\startdata
A1 & $0.163\pm0.003$ & $-1.428\pm0.003$ & -- & -- & -- & $15.64\pm0.01$ \\
A2 & $0.416\pm0.003$ & $-1.214\pm0.003$ & -- & -- & -- & $16.09\pm0.01$ \\
B & $\equiv 0.000$ & $\equiv 0.000$ & -- & -- & -- & $17.11\pm0.01$ \\
C & $-0.572\pm0.003$ & $-1.042\pm0.003$ & -- & -- & -- & $17.33\pm0.02$ \\
G & $-0.074\pm0.012$ & $-0.798\pm0.008$ & $0.47\pm{0.36}$ & $0.35\pm{0.21}$ & $53\pm{42}$ & $18.80\pm0.43$ \\
\enddata
\end{deluxetable*}

\begin{figure}[t]
\includegraphics[width=0.5\textwidth]{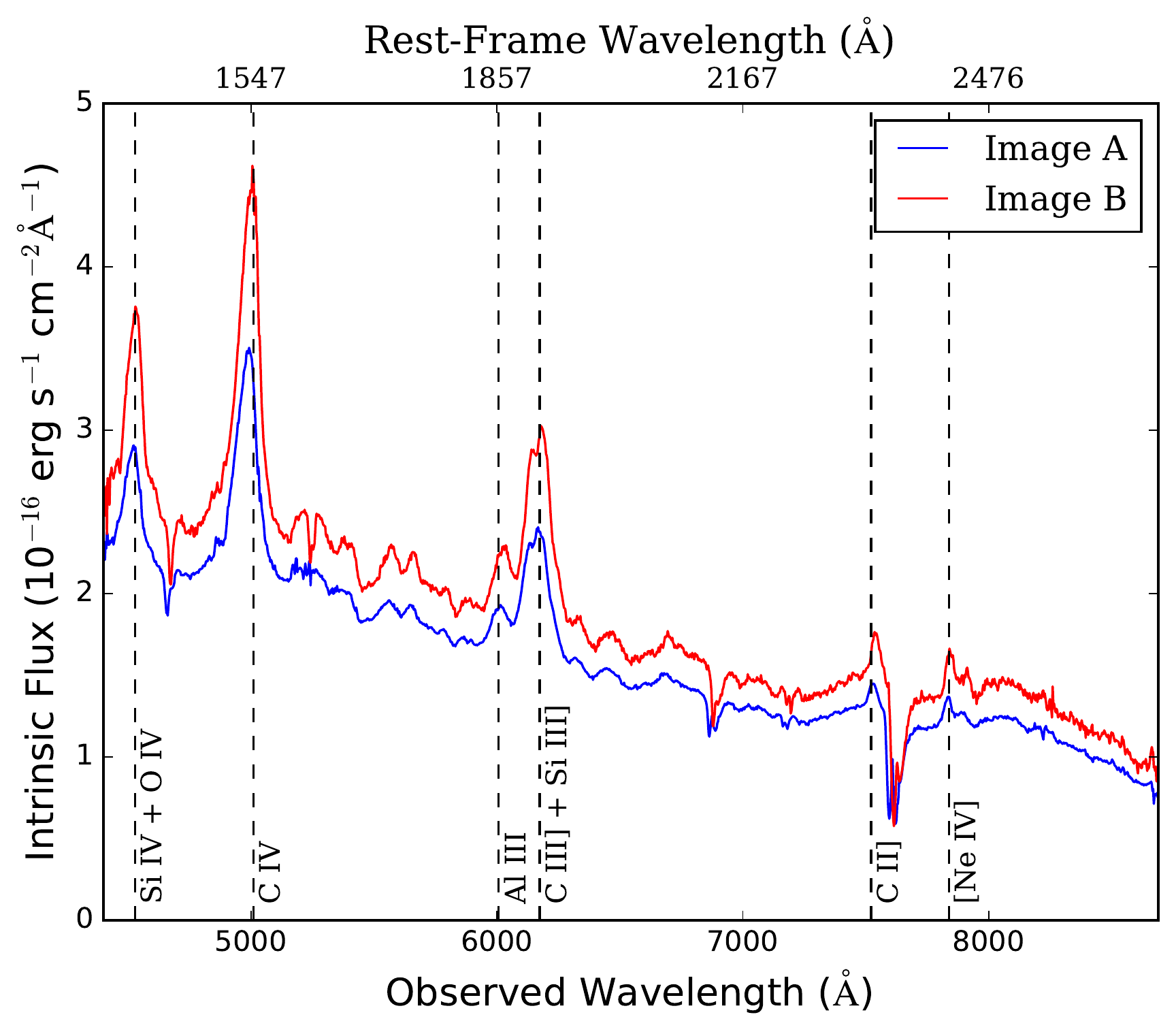}
\caption{Reduced VLT FORS1 spectra for images A (blue) and B (red), magnification corrected to intrinsic flux estimates.  The locations of several known emission lines are marked and the corresponding rest-frame wavelength is indicated along the top of the plot.}
\label{fig:spectra}
\end{figure}

The angular separation between images in WFI2026 is small, with a scale size of ${\sim}\,1\farcs4$ \citep{morg2004a, vero2010a}, making photometric measurements challenging.  Nevertheless with a typical seeing of $1\farcs6$ with C2 and $1\farcs4$ with ECAM, the angular separations are above the Nyquist limit ($\approx \frac{1}{2}\,\text{seeing} = 0\farcs8$) for most image pairs.  Images B and C are separated from each other and A1 and A2 by at least $1\farcs0$. The merging pair A1 and A2, however, are only separated by ${\sim}\,0\farcs3$, too close to resolve fluxes in the individual images.  We reduced the data and performed our subsequent analysis using the combined flux A=A1+A2.  We employed the Magain, Courbin, and Sohy (MCS) deconvolution algorithm of \citet{maga1998a, maga2007a} for de-blending flux from the multiple images.  Using a point spread function (PSF) measured from nearby reference stars, this algorithm computes a high-resolution deconvolved image of the quasar.  We then followed the approach discussed in \citet{vuis2007a, vuis2008a} to find flux in individual images, including priors on astrometry from \citet{chan2010a} to further improve accuracy.   We measured image fluxes in each subexposure and calculated the median value at each epoch to provide the reduced lightcurves shown in Figure~\ref{fig:lightcurves} and in Appendix~\ref{sec:lc_table}.

The compact angular size of the system gave rise to significant cross-talk between photometric measurements.  This additional noise, in excess of photon shot-noise, is typical for multiply-lensed quasars.  For image C in WFI2026, however, this cross-talk noise was on the same order as the intra-season variability and could mimic a microlensing signal.  To mitigate this effect, we averaged over each season for image C and used these season averages in our microlensing analysis. 
We determined each season average for image C, $\langle c\rangle$, using least-squares fitting.  The error bars, $\langle \sigma_{c}\rangle$, were selected such that within each season the $\chi^2$ value, relative to the reduced lightcurve, was equal to the number of exposures, $N$, for the season.  This amounted to numerically solving 
\begin{equation}
N = \sum_i^N \frac{(c_i - \langle c\rangle)^2}{\sigma_i^2 + \langle\sigma_{c}\rangle^2}.
\label{eqn:seas_avg}
\end{equation}
for $\langle\sigma_{c}\rangle$.
The values $c_i$ and $\sigma_i$ came from the reduced light curve.  This simplification removed our ability to discern intra-season variability, but allowed us to more confidently measure the annual variability, which dominates the microlensing signal in WFI2026 \citep{mosq2011a}.  We show the image C season averages overlaid atop the reduced lightcurve in Figure~\ref{fig:lightcurves}.

For lens modeling we used \textsl{HST} imaging from the CfA-Arizona Space Telescope Survey (CASTLES\footnote{https://www.cfa.harvard.edu/castles/}) \citep{muno1998a, koch1999a, leha2000a}.  Our exposure was taken on October 21, 2003 with the Near-Infrared Camera and Multi-Object Spectrograph (NICMOS) through the F160W filter \citep{morg2004a}.  From this image, we derived astrometry using the \textit{imfitfits} routine of \citet{leha2000a}.  Our astrometric fits, including the shape parameters for a de Vaucouleurs lens galaxy, are shown in Table~\ref{tab:hst_astrom}.  Because the lens galaxy in WFI2026 is faint, the galaxy model parameters have sizable uncertainties, but the results are in agreement with values reported in \citet{chan2010a}.

We also analyzed spectra obtained using the Very Large Telescope (VLT) of the European Southern Observatory (ESO) with the FORS1 mult-object spectrograph.   In our analysis we use a series of fourteen $1400\,s$ exposures through the GG435 filter obtained between 2004 and 2006.   The slit was oriented to capture both images A (A1 + A2) and B.  Each exposure spanned the observed wavelength range of $4400\text{--}8690\,\text{\AA}$. 
For WFI2026 at $z_s=2.23$ we fully resolved the \ion{C}{4} line which allowed us to estimate the black hole mass (see section~\ref{sec:mbh}).  We display these spectra in Figure~\ref{fig:spectra}.

\section{Macro Lens Modeling}
\label{sec:lens_model}

Our microlensing analysis required a model of the strong (macro) lensing in the system.   WFI2026 has been previously modeled in \citet{slus2012a} and \citet{bate2018a}.  In both cases, the authors found the need for significant external shear, but achieved good fits using a singular isothermal ellipsoid with shear (SIE+$\gamma$) model.  From the isothermal ellipsoid models of \citet{slus2012a} and \citet{bate2018a} we estimated the velocity dispersion in the lens. 

Following the convention of previous multi-epoch studies \citep[e.g.][]{morg2018a}, we also modeled the lens with a sequence of two-component de Vaucouleurs and Navarro, Frenk, and White (NFW) \citep{nava1997a} models with external shear.  This simulated the expected profiles of stellar matter (de Vaucouleurs) and dark matter (NFW) and allowed us to marginalize over the unknown dark matter fraction.  We used the \texttt{LENSMODEL} software \citep{keet2001a}, omitting constraints from the flux ratios, which can be influenced by microlensing.  Our first model employed a pure de Vaucouleurs profile (100\% stellar matter) to fit for the centroid, moment, ellipticity, position angle,  effective radius, and shear strength and orientation of the lens galaxy.
We call this model $f_{M/L}=1.0$, where we define $f_{M/L}$ as the fractional strength of the de Vaucouleurs moment relative to a unity mass-to-light-ratio model.
We then reduced the de Vaucouleurs moment in increments of 10\% and added an NFW component fixed to the same lens centroid, ellipticity, and position angle, and re-fit for all parameters.  This generated a ten-model sequence with $f_{M/L} = 0.10\text{--}1.0$ in increments of 0.1, nominally spanning the range of 0-90\% dark matter, and permitted our analysis to marginalize over the unknown dark matter fraction. 
An advantage of using a wide range of dark matter fractions is that it also effectively samples a wide range of possible lens galaxy profile shapes.  This is especially important for WFI2026 given the broad errors in Table~\ref{tab:hst_astrom}, which can predict very different stellar mass fractions.
Best fits for the convergence, $\kappa$, and shear, $\gamma$, and stellar-to-total-convergence ratio, $\kappa_*/\kappa$, of this sequence are given in Table~\ref{tab:lens_models}.   We also show the relative quality of fit, which varies little between models.  

A shortcoming of our WFI2026 lens model sequence is the unknown lens redshift, $z_{l}$.  In the discovery paper, \citet{morg2004a} favor a lens redshift of $z_{l}=0.4$, based upon lens galaxy luminosity.  However, in a more recent study \citet{mosq2011a} used astrometry-based methods \citep{ofek2003a} to estimate a lens redshift of $z_{l}=1.04$.  As our VLT spectra showed no apparent lens galaxy features, we were unable to independently measure the redshift.  We note that the lack of any lens galaxy signal in the spectrum is consistent with a featureless UV continuum of a $z_{l}=1.04$ or greater elliptical galaxy.  
In our analysis we adopted the more recent estimate of $z_{l}=1.04$ as the nominal lens redshift, but we also examined the impact of instead using $z_{l}=0.4$ in section~\ref{sec:ml_results}.

\begin{deluxetable*}{c|cccc|cccc|cccc|c}
\tablecaption{Convergence, shear, and stellar convergence fraction $\kappa_{*}/\kappa$ for each lens model.  The parameter $f_{M/L}$ indicates the strength of the de Vaucouleurs moment relative to a de Vaucouleurs-only model, providing a proxy for the luminous matter fraction.  The values for $\kappa$, $\gamma$, and $\kappa_{*}/\kappa$ are calculated at the position of each image.  The final column, $\chi^2/N_{dof}$, indicates the relative fit of the model across all images.\label{tab:lens_models}}
\tablehead{
\multirow{2}{0.05\textwidth}{$f_{M/L}$} & \multicolumn{4}{c|}{Convergence $\kappa$} & \multicolumn{4}{c|}{Shear $\gamma$} & \multicolumn{4}{c|}{$\kappa_{*}/\kappa$} & \multirow{2}{0.05\textwidth}{$\chi^2/N_{\text{dof}}$} \\ 
\colhead{} \vline &\colhead{A1} & \colhead{A2} & \colhead{B} & \colhead{C} \vline & \colhead{A1} & \colhead{A2} & \colhead{B} & \colhead{C}  \vline& \colhead{A1} & \colhead{A2} & \colhead{B} & \colhead{C} \vline & \colhead{}
 }
\startdata
0.1 & 0.90 & 0.91& 0.88& 0.91 & 0.08 & 0.11& 0.06& 0.14 & 0.04 & 0.05& 0.02& 0.05 &  3.60 \\
0.2 & 0.85 & 0.86& 0.83& 0.86 & 0.12 & 0.16& 0.09& 0.20 & 0.04 & 0.05& 0.02& 0.05 &  2.85 \\
0.3 & 0.76 & 0.78& 0.72& 0.78 & 0.20 & 0.27& 0.14& 0.34 & 0.11 & 0.13& 0.07& 0.13 &  3.10 \\
0.4 & 0.70 & 0.71& 0.65& 0.70 & 0.26 & 0.34& 0.18& 0.42 & 0.15 & 0.16& 0.10& 0.16 &  3.16 \\
0.5 & 0.63 & 0.65& 0.57& 0.64 & 0.31 & 0.42& 0.21& 0.52 & 0.20 & 0.22& 0.13& 0.21 &  2.82 \\
0.6 & 0.56 & 0.58& 0.50& 0.57 & 0.37 & 0.49& 0.25& 0.61 & 0.26 & 0.28& 0.18& 0.27 &  2.89 \\
0.7 & 0.51 & 0.53& 0.44& 0.52 & 0.41 & 0.56& 0.28& 0.69 & 0.31 & 0.34& 0.21& 0.33 &  2.77 \\
0.8 & 0.40 & 0.43& 0.31& 0.42 & 0.50 & 0.68& 0.35& 0.84 & 0.56 & 0.59& 0.43& 0.58 &  2.89 \\
0.9 & 0.33 & 0.36& 0.23& 0.35 & 0.56 & 0.76& 0.38& 0.94 & 0.75 & 0.77& 0.64& 0.77 &  2.87 \\
1.0 & 0.26 & 0.30& 0.16& 0.29 & 0.61 & 0.83& 0.42& 1.03 & 1.00 & 1.00& 1.00& 1.00 &  2.84 \\
\enddata
\end{deluxetable*}

\begin{deluxetable}{c|c|c}
\tablecaption{Time delays used for the microlensing analysis.\label{tab:time_delays}}
\tablehead{
\colhead{Source} & \colhead{$\tau_{B-A}$ (days)} & \colhead{$\tau_{B-C}$ (days)}
}
\startdata
Euler ECAM & $18.7 ^{+4.1}_{-4.3}$ & -- \\
Lens Models & -- & $23.7^{+5.2}_{-5.2}$ \\
\enddata
\end{deluxetable}

We attempted to measure the time delays using the \texttt{PyCS} algorithm of \citep{tewe2013a}, which performed very well in a recent time-delay challenge \citep{liao2015a, bonv2016a}.   \texttt{PyCS} has been adopted as the curve-shifting algorithm of choice for COSMOGRAIL.  Details of the WFI2026 measurement will be included as part of a larger set of time delay measurements in a forthcoming paper \citep[in prep]{mill2019a}.  Here we report only the resulting time delays, shown in Table~\ref{tab:time_delays}.  The time delays are consistent with models at $z=1.04$ but not at $z=0.4$, lending further support to the larger lens redshift.
The empirical time delays for image C were highly uncertain, and inconsistent between A-C and B-C.  As such, we retained only the A-B measurement in our microlensing analysis.  For image C, we instead used the $z=1.04$ lens model that best matched the empirical A-B delay, and extrapolated the model results to estimate a time delay.  
We explored the impact of time delay uncertainty on our disk size measurement in section~\ref{sec:ml_results}.

\section{Microlensing Models}
\label{sec:ml_model}

Our microlensing analysis was based on the procedure developed in \cite{koch2004a} and \citet{koch2006a}.  This technique uses Monte Carlo methods to fit the observed microlensing lightcurves from trajectories through a set of stellar magnification fields.

Before running our analysis, we binned the lightcurves in a 20-day window, using error-weighted mean magnitude and mean Heliocentric Julian Date (HJD).  This decreased the number of epochs from 548 to 129 which kept calculation times reasonable for the subsequent Monte Carlo analysis. Because
\citet{mosq2011a} estimated a source-crossing timescale of $1.4$ years in WFI2026 and a longer Einstein-radius-crossing timescale of $26.6$ years we were not concerned about microlensing on a sub-monthly scale.  These short time-scales are also not well-resolved for typical trajectories across the microlensing patterns.
To further mitigate the impact of short-timescale noise on the microlensing solution, we included systematic errors of $0.015\,\text{mag}$ to account for any unmodeled photometric errors.

\begin{figure*}[t]
\includegraphics[width=\textwidth]{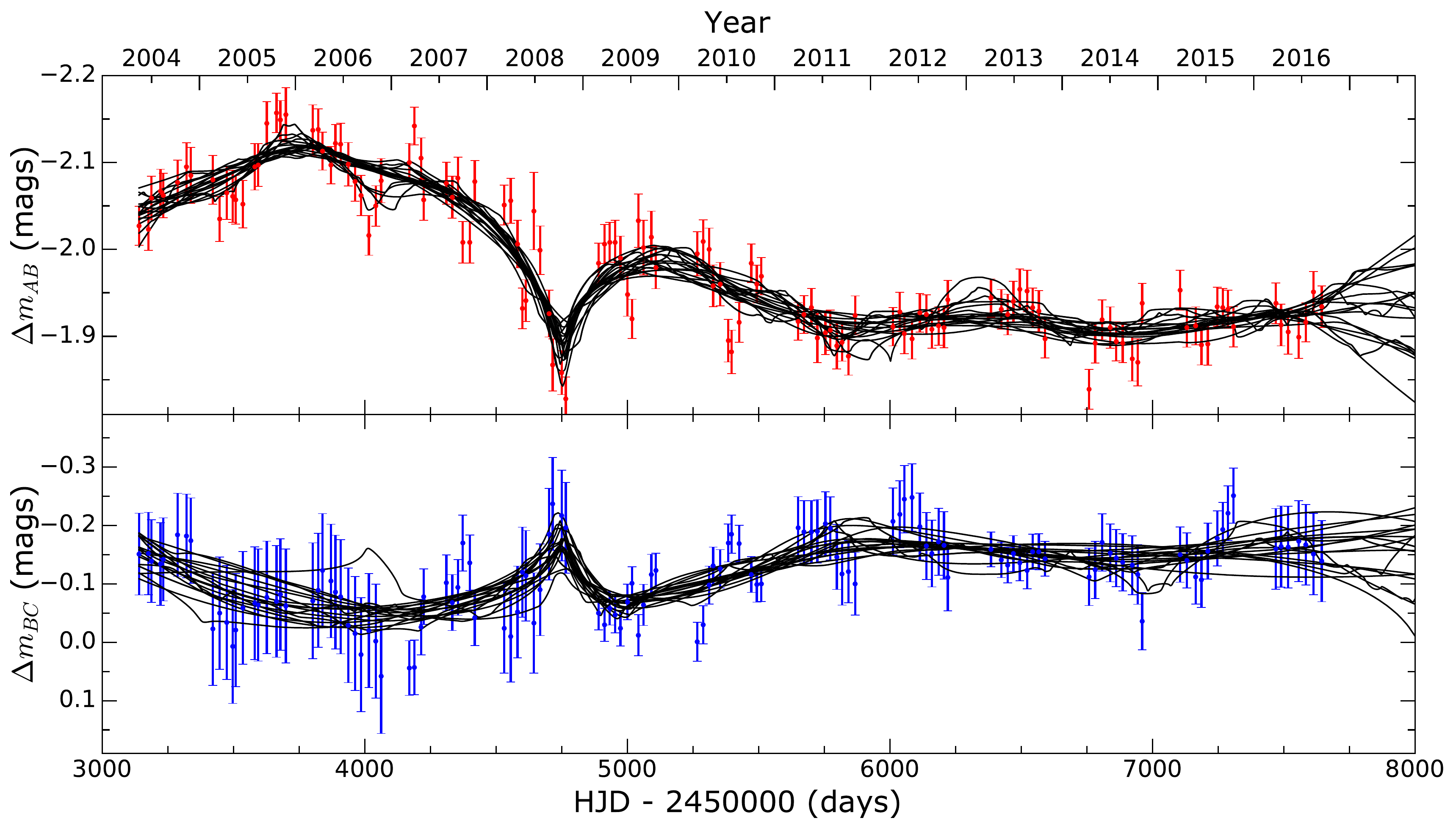}
\caption{The 20 best-fit curves from our microlensing analysis.  \textit{Top}: Time-delay corrected difference curves for images A-B, $\Delta m_{AB} = m_A - m_B$, in magnitudes.  \textit{Bottom}: The difference curves for images B-C, $\Delta m_{BC} = m_B - \langle m_C \rangle$ where $\langle m_C \rangle$ is the season average for image C.  All curves fit the strong microlensing feature from image B near HJD-2450000 = 4700 days.  They also consistently fit the slow gradient between HJD-2450000 ${\sim}\,3500 \text{--} 6000$ days.}
\label{fig:good_ml_fits}
\end{figure*}

We shifted each curve by the time delays, holding the lightcurve for B as a fixed reference and linearly interpolating for images A and C.  We averaged over the time-delay shifted season C values as detailed in section~\ref{sec:data}.  The magnitude differences are shown in Figure~\ref{fig:good_ml_fits} for B-A (top panel, red) and B-C (bottom panel, blue).
Microlensing is the dominant source of time variability between these time-delay-shifted difference lightcurves. 


In a dynamic microlensing analysis, the magnification curve, $\mu(t)$, depends on highly nonlinear magnification by a stellar field in the lens galaxy.  Because we cannot measure the precise stellar characteristics in the lens galaxy, we generated many possible magnification patterns at the range of $\kappa_*/\kappa$ from our model sequence (see Table~\ref{tab:lens_models}).  As with previous studies, we assumed a stellar mass distribution of $dN/dM \propto M^{-1.3}$, a ratio of maximum over minimum mass of 50, and a variable median microlensing mass $\langle M_*/M_{\odot} \rangle$.  We projected the stellar magnification patterns on a $8192\times 8192$ grid which spanned sizes from $40 R_{\text{E}}$ down to the pixel scale, ${\sim}\,0.005 R_{\text{E}}$.  Here $R_{\text{E}}=D_{OS}\theta_{\text{E}}$, where $D_{OS}$ is the angular diameter distance from the observer to the source and $\theta_{\text{E}} \propto \langle M_{*}/M_{\odot} \rangle^{1/2}$ is the Einstein radius as a function of median stellar microlens mass.  

For each of the ten macro models ($0.1 \leq f_{M/L} \leq 1.0$), we generated 40 magnification patterns for each lensed image, yielding 400 complete sets of magnification patterns.  This eliminated concerns about the introduction of systematics from repetitive use of one or a small number of patterns for a given image location and macro model.  We created fits to the combined image A = A1+A2 light curve by summing the model light curves generated separately from the magnification patterns for images A1 and A2. The goodness-of-fit to each point in the summed light curve A was assigned the same statistical weight as for each point in the resolved light curves for images B and C.

To model the disk radius, we convolved each magnification pattern with a Gaussian kernel at a range of trial accretion disk sizes.  We chose seventeen radii evenly spaced in the logarithmic range $\log(r/\text{cm}) = 14.5-18.5$.  Since \citet{mort2005a} showed that the half light radius, rather than profile shape, affects the inferred microlensing size, we selected the Gaussian profile rather than the thin disk model for speed of calculation.   
Upon conclusion of the analysis, we converted the best-fit Gaussian scale radius to a thin disk scale radius $r_s$.   

We generated trial lightcurves by moving a point source across a convolved magnification pattern.  We selected transverse velocities from the logarithmic range $10\,\text{km s$^{-1}$} \leq \hat{v_e}\langle M_*/M_{\odot} \rangle^{1/2} \leq 10^6\,\text{km s$^{-1}$}$ and randomized the directions and starting points in each image.  From these trajectories, we found the magnification as a function of time and compared this to our empirical lightcurves using a $\chi^2$ statistic.  We allowed for a $0.5\,\text{mag}$ systematic uncertainty in the intrinsic flux ratios between the images to account for the influence of substructure and broad line region contamination.  We terminated a trial when $\chi^2/N_{\text{dof}} > 1.4$, where $N_{\text{dof}}$ is the number of degrees of freedom, as these solutions did not contribute significant statistical weight to the inferred parameter values.  In the Monte Carlo phase of our analysis, we used the United States Naval Academy High Performance Cluster\footnote{https://www.usna.edu/ARCS/} to attempt $10^7$ trials on each of the 400 magnification patterns for a grand total of $4\times10^9$ trials.  Additional trials did not significantly improve constraints. 
\section{Results}
\label{sec:results}

In this section we present our microlensing analysis and show our determination of the size of WFI2026, reported as the scale radius $r_s$.  We also present our analysis of the VLT spectra to determine the mass of the black hole.  

\begin{figure*}[t]
\gridline{\fig{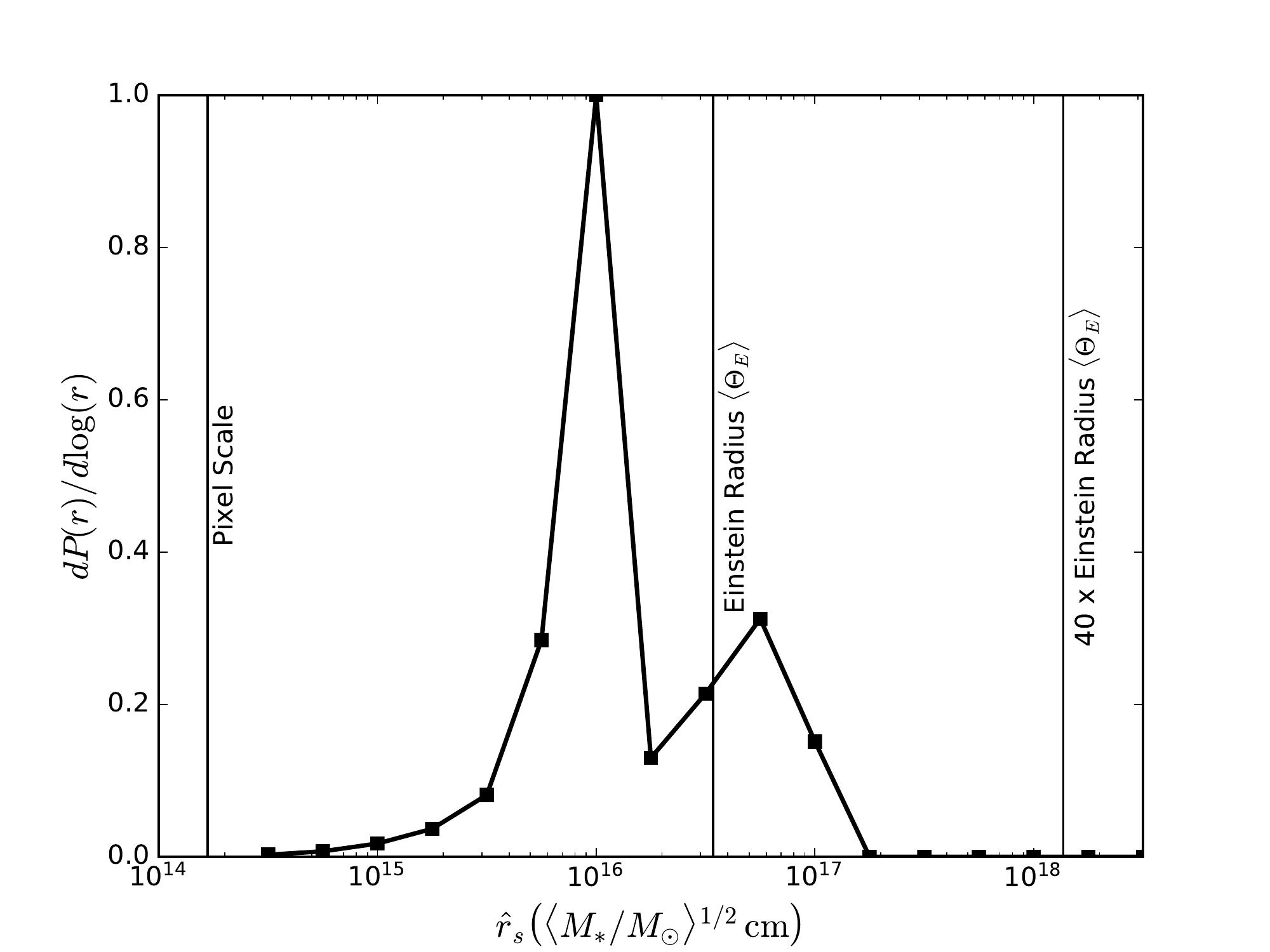}{0.5 \textwidth}{}
             \fig{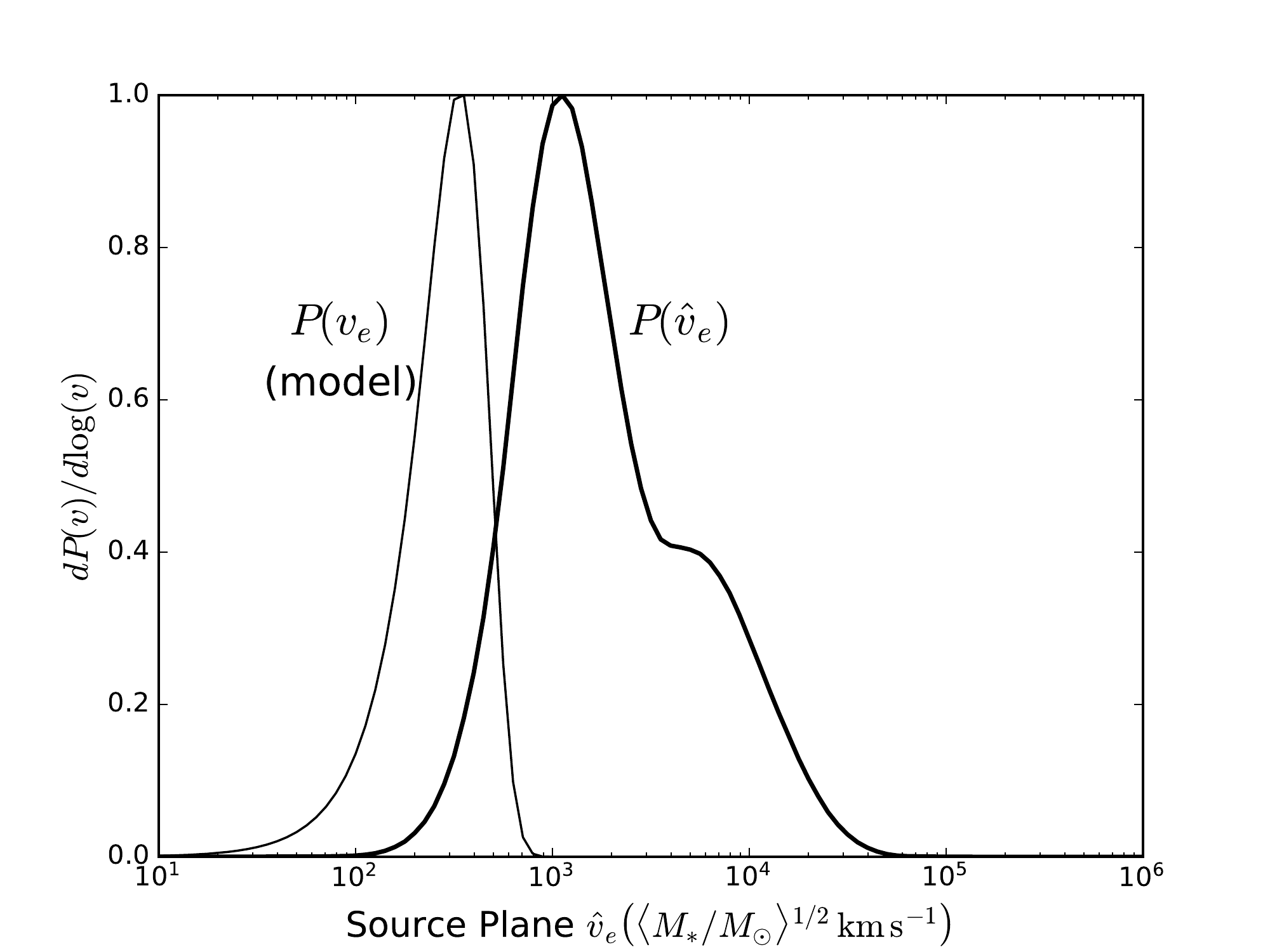}{0.5 \textwidth}{}}
\caption{Effective source radius and velocity inferred from the WFI2026 microlensing analysis.  \textit{Left:} Probability density for the scale radius, $\hat{r}_s$ in Einstein units, scaled to a $1\,M_{\odot}$ median microlensing mass.  \textit{Right:}  The probability density function for the effective quasar velocity, $\hat{v}_e$ in Einstein units, scaled to a $1\,M_{\odot}$ median microlensing mass.   The thicker curve indicates the value found from the microlensing analysis and shows the same bimodality as seen in the probability density for the radius.  The thin curve indicates our cosmological velocity model, $v_e$, in $\text{km}\,\text{s}^{-1}$. }
\label{fig:vhat}
\end{figure*}

\begin{figure*}[t]
\centering
\includegraphics[width=0.5\textwidth]{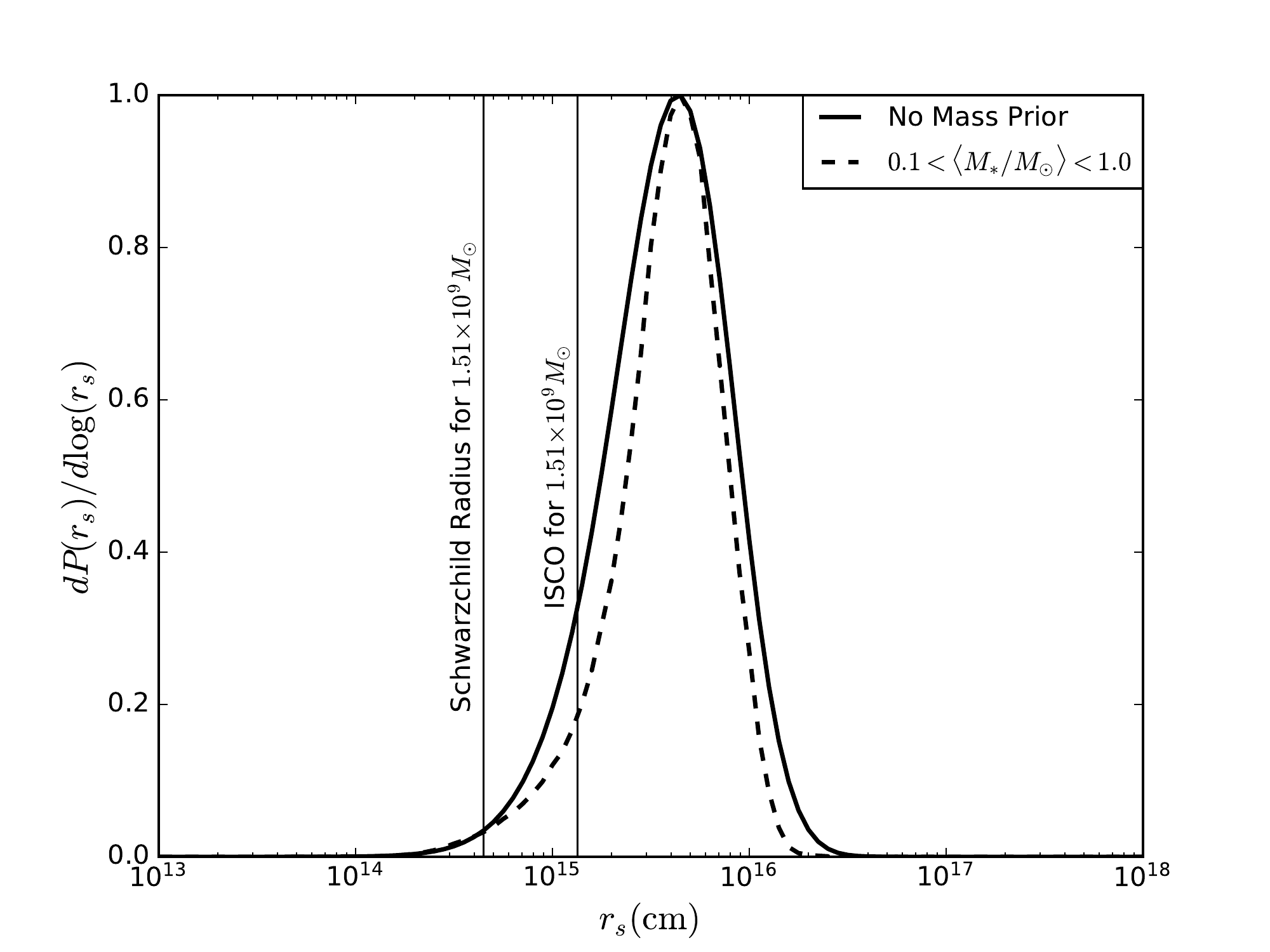}
\caption{Accretion disk radius of WFI2026 $\lambda_{\text{rest}} = 2043\AA$, displayed as a probability density for the scale radius in physical units.  The solid line corresponds to the result with no prior on median microlensing mass.  The dashed line is the solution with a uniform prior on the median microlensing mass of $0.1 < \langle M_{*}/M_{\odot}\rangle < 1.0$.}
\label{fig:rhat}
\end{figure*}

\subsection{Microlensing}
\label{sec:ml_results}

\begin{figure}[t]
\centering
\includegraphics[width=0.5\textwidth]{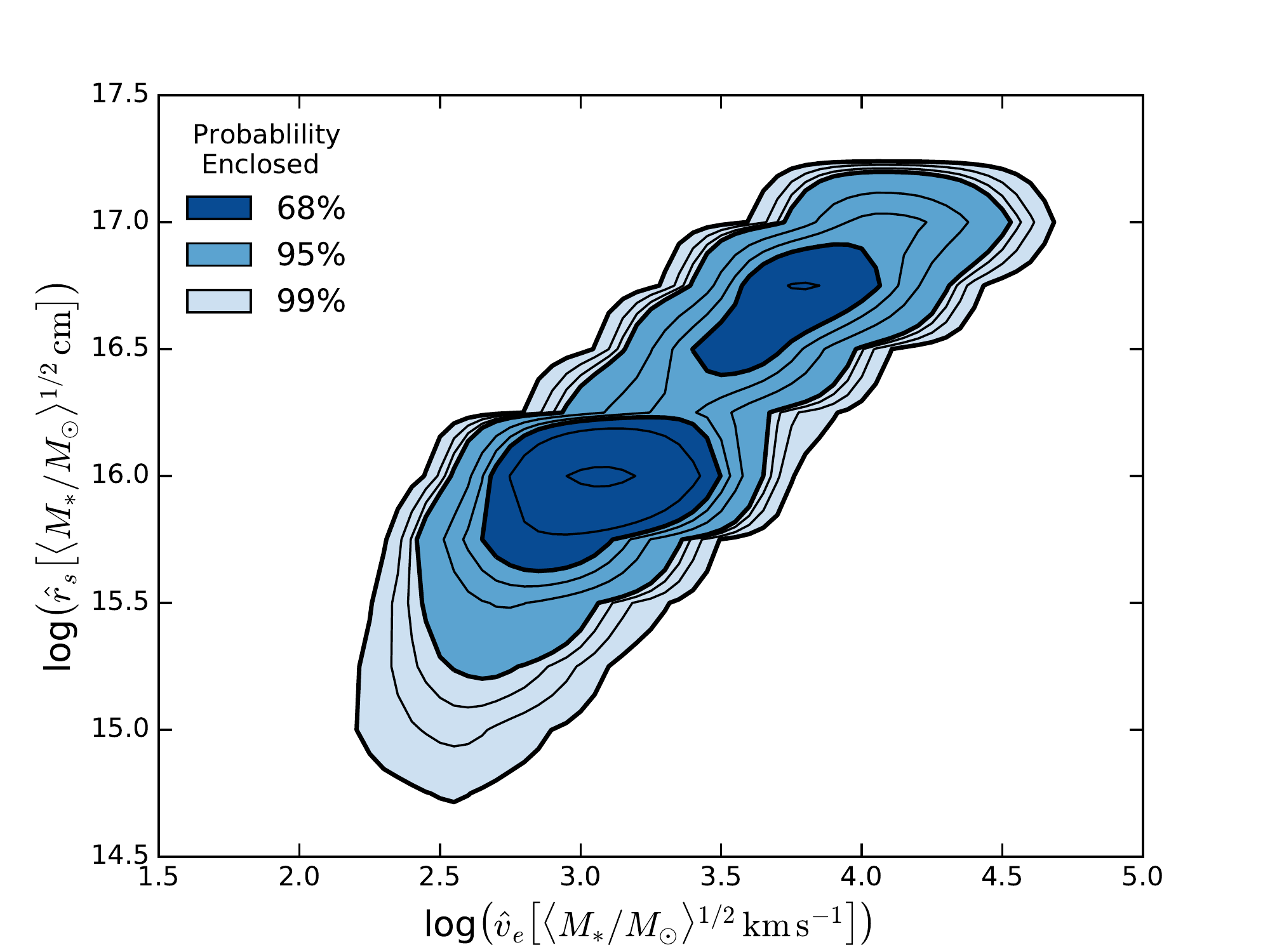}
\caption{Probability contours of $\log(\hat{v}_e)$ vs. $\log(\hat{r}_s)$ for WFI2026.  Confidence intervals enclosing 68\%, 95\%, and 99\% of the total probability are shaded in blue.  The bimodal peaks are evident here, as is an overall linear relation between $\log(\hat{v}_e)$ and $\log(\hat{r}_s)$.}
\label{fig:rhat_vhat}
\end{figure}

Several of the best-fits to the time-delay corrected curves are shown in Figure~\ref{fig:good_ml_fits}.   We can see strong microlensing variability in this system on the order of ${\sim}\,0.2\,\text{mag}$ over the 13 seasons.  There is also a short duration, high-magnification event in image B rising across the entire 2008 season.  The best-fit curves consistently reproduced these dominant microlensing features.

We calculated probability densities for the variables of interest by marginalizing over the other variables of the model.  For the radius, this took the form
\begin{equation}
P(\hat{r}_s|D) \propto \int_{0}^{\infty} P(D|\hat{r}_s, \xi)\pi(\xi)\pi(\hat{r}_s)d\xi.
\label{eqn:marginalize}
\end{equation}
Here $\xi$ represents all other variables, including effective source velocity, $\hat{v}_e$ and luminous matter fraction, $f_{M/L}$.  The prior distribution is captured in $\pi(\xi)$ and is, for example, log-uniform for $\hat{v}_e$ on $[10, 10^6]$ and uniform for $f_{M/L}$ on $[0.1, 1]$ while the prior $\pi(\hat{r}_s)$ is log-uniform on $[10^{14.5}, 10^{18.5}]$.  The probability of the data $P(D|\hat{r}_s, \xi)$ is equivalent to $P(\chi^2|N_{\text{dof}})$ in equation 10 of \citet{koch2004a}. 

In Figure~\ref{fig:vhat}, we display the resulting probability density for the primary variable of interest, the source size $\hat{r}_s=r_s\langle M_*/M_{\odot} \rangle ^{-1/2}$.  The $\hat{r}_s$ distribution is in Einstein units, scaled assuming a $1\,M_{\odot}$ median stellar mass in the lens galaxy.  To convert this result to physical units, we convolve $\hat{r}_s$ with the probability density for $\langle M_*/M_{\odot} \rangle$, where the $\langle M_*/M_{\odot} \rangle$ distribution is found as in \citet{koch2004a} by
\begin{equation}
P(\langle M_*/M_{\odot} \rangle|\text{D}) \propto \int P(\hat{v}_e|\text{D}) P(v_e)dv_e
\label{eqn:mhat}
\end{equation}
where $\hat{v_e}=v_e\langle M_*/M_{\odot} \rangle^{-1/2}$.
The velocity probability density distributions are shown in the right panel of Figure~\ref{fig:vhat}.  Because $\hat{v}_e$ is more finely sampled than $\hat{r}_s$, the resulting distribution is much smoother.

To find $P(v_e)$, the probability density for the actual source velocity, we model the effective source velocity $v_e$ following the method of \citet{koch2004a} and using the formulation of \citet{mosq2011a}.
This includes velocity contributions from the observer, $v_{\text{CMB}}$, source, $\sigma_{\text{pec}}(z_s)$, lens bulk motion, $\sigma_{\text{pec}}(z_l)$, and lens velocity dispersion, $\sigma_*$.   
We projected the CMB dipole along the line of sight to WFI2026 to find the north and east vector components of $v_{\text{CMB}}$ as $-227\,\text{km}\,\text{s}^{-1}$ and $-244\,\text{km}\,\text{s}^{-1}$ respectively.
 For the bulk galaxy motions we used cosmological models to estimate the one-dimensional peculiar velocity dispersions to be $\sigma_{\text{pec}}(z_l) = 265\,\text{km}\,\text{s}^{-1}$ and $\sigma_{\text{pec}}(z_s)=204\,\text{km}\,\text{s}^{-1}$.  We also included a contribution of $\sigma_* = 335\,\text{km}\,\text{s}^{-1}$ from the stellar velocity dispersion in the lens, determined from the Einstein radius found by \citet{slus2012a} with an SIE+$\gamma$ lens model.  
 
With our inferred distribution for $\langle M_*/M_{\odot} \rangle$ we estimated the probability density function for the accretion disk size in physical units, $r_s$, shown with the solid line in Figure~\ref{fig:rhat}. 
Adopting a nominal inclination of $\langle \cos(i)\rangle = 0.5$ for ready comparison to other microlensing studies \citep[e.g.][]{blac2011a, morg2018a}, the scale radius of the WFI2026 accretion disk at $\lambda_{\text{rest}} = 2043\,\text{\AA}$ is $\log\{(r_s/\text{cm}) [\cos(i)/0.5]^{-1/2}\} = 15.74^{+0.34}_{-0.29}$.   This result can be easily re-scaled to any inclination, such as a smaller angle of $\la30\degr$ that may be more typical of quasars under unification models \citep[e.g.][]{beve1995a, urry1995a}.

To examine the impact of uncertainty in the median microlensing mass distribution, we also experimented with applying a uniform mass prior of $0.1 < \langle M_*/M_{\odot} \rangle < 1.0$, resulting in the distribution with the dotted line in Figure~\ref{fig:rhat}.  The mass prior narrows the distribution marginally, but nonetheless provides a fully consistent result.  For self-consistency, we adopt the distribution without the mass prior as our primary result.
 
We found a bimodal distribution in both $dP(\hat{r}_s)/d\log(\hat{r}_s)$ and $dP(\hat{v}_e)/d\log(\hat{v}_e)$ but see no evidence for bimodality in $dP(r_s)/d\log(r_s)$.  
We can understand this by examining the underlying nature of the bimodal solutions.  The high-velocity, large-radius mode corresponds to a low median microlensing mass.
However, because the median microlensing mass is smaller for these solutions, the true value of $r_s = \hat{r}_s \langle M_*/M_{\odot}\rangle^{1/2}$, the product of mass and radius, remains relatively unchanged.  Both modes return the same physical estimate in the limit of $\hat{v}_e \propto \hat{r}_s$.  This relation is nearly satisfied by our solutions as seen in the probability contours in Figure~\ref{fig:rhat_vhat}.  This trend indicates that the data most strongly constrain the ratio of $\hat{v}_e/\hat{r}_s$, which is independent of $\langle M_*/M_{\odot}\rangle$.  This insensitivity to the unknown microlensing mass is one of the strengths of multi-epoch lightcurve analysis.

As an additional verification, we re-ran the microlensing analysis with a log-uniform prior on the source velocity of $1.0 < \log\hat{v}_e < 4.0$.  This disallowed unreasonably high-velocity solutions, providing a different means of imposing a lower limit on microlensing mass.  In this second analysis, the velocity distribution had only a single mode, as expected, and the resulting size estimate was effectively unchanged.

We also tested the sensitivity of our results to the unknown lens redshift, $z_{l}$, by calculating the distances, velocities, and radii for the low and high estimates of the lens redshift, $0.4 < z_{l} < 1.04$.  Given the lack of a caustic crossing in the WFI2026 lightcurves, the dominant timescale for microlensing variability is the Einstein radius crossing time $t_{E} = R_{E}/v_e$.  Because the physical radius $R_{E} \propto (D_{\text{LS}}D_{\text{OS}}/D_{\text{OL}})^{1/2}$, the influence of the resulting sizes only scales as the square root of the change in the angular diameter distances.  Comparing the physical size measurement between the two redshifts we found a change in $\log(r)$ of less than 2\%, fully consistent within the statistical errors.

Similarly, our results are only weakly sensitive to the time delay.  We repeated the microlensing analysis with low and high time delays based on the error limits in Table~\ref{tab:time_delays}, but the resulting changes in our measurement of the scale radius were negligible.

We attempted to estimate the relative stellar mass fraction by marginalizing over velocities and radii.
There was a mild preference for the lowest stellar mass models, but not significant enough to warrant a quantitative estimate.

\subsection{Black Hole Mass}
\label{sec:mbh}

Our reduced VLT spectra for images A and B are shown in Figure~\ref{fig:spectra} with several prominent emission lines indicated.  Because of the relatively high source redshift, $z_s = 2.23$, the \ion{Mg}{2} line is redshifted out of the observed frame.  The \ion{C}{4} line is, however, fully resolved in both images and can be used to estimate black hole mass \citep{vest2006a, park2013a}.

We first estimated the \ion{C}{4} emission line width in the spectra
  of image A and image B independently.  Line widths were measured by
  fitting a local, linear continuum under the emission line and then
  determining the full width at half maximum (FWHM) and the line
  dispersion or second moment, $\sigma_l$, directly from the data
  above the continuum (see \citet{pete2004a} for a more detailed
  description).  There is good agreement in the line widths determined
  from the spectra of the two separate images, with average values of
  FWHM$=6015\pm40$\,km\,s$^{-1}$ and
  $\sigma_l=3616\pm4$\,km\,s$^{-1}$.

\citet{vest2006a} and \citet{park2013a} provide
prescriptions for estimating black hole masses based on the
\ion{C}{4} emission line that are calibrated to the H$\beta$
  reverberation mapping results for local AGNs.  Work by \citet{denn2013a} shows that single-epoch black hole masses derived from
  the \ion{C}{4} emission line are less biased when adopting
    $\sigma_l$ as the line width measurement rather than FWHM, so we
    focus on those prescriptions.  The other necessary ingredient is
    the continuum luminosity at rest-frame 1350\,\AA, which was not
    covered in the VLT spectra of WFI 2026-4536.  Fortunately,
    \citet{vest2006a} show that $L_{\lambda}$(1450\,\AA)
    can be directly substituted for $L_{\lambda}$(1350\,\AA).

We measured continuum flux densities of
$f_{\lambda}$($1450\times(1+z)$)$=1.3\times10^{-15}$\,erg\,s$^{-1}$\,cm$^{-2}$\,\AA$^{-1}$
for image B and
$6.9\times10^{-15}$\,erg\,s$^{-1}$\,cm$^{-2}$\,\AA$^{-1}$ for image A.
Allowing for the range of image magnifications spanned by the model
sequence in Table~\ref{tab:lens_models}, and accounting for Galactic extinction along the
line of sight, we find $\log(\lambda L_{\lambda}$(1450\,\AA)$\bm{/\text{ergs}\,\text{s}^{-1}})=46.523^{+0.388}_{-0.155}$.
Combined with the emission line width and using the prescriptions of
\citet{vest2006a} we estimate $\log(M_{\rm BH}/M_{\odot})
= 9.18^{+0.39}_{-0.34}$, including in our uncertainty the 0.33\,dex of
scatter reported for their prescription.  The black hole mass estimate
is nearly identical if the prescriptions of \citet{park2013a} are
adopted instead.

\section{Discussion and Conclusions}
\label{sec:discussion}

\begin{figure}[t]
\centering
\includegraphics[width=0.5\textwidth]{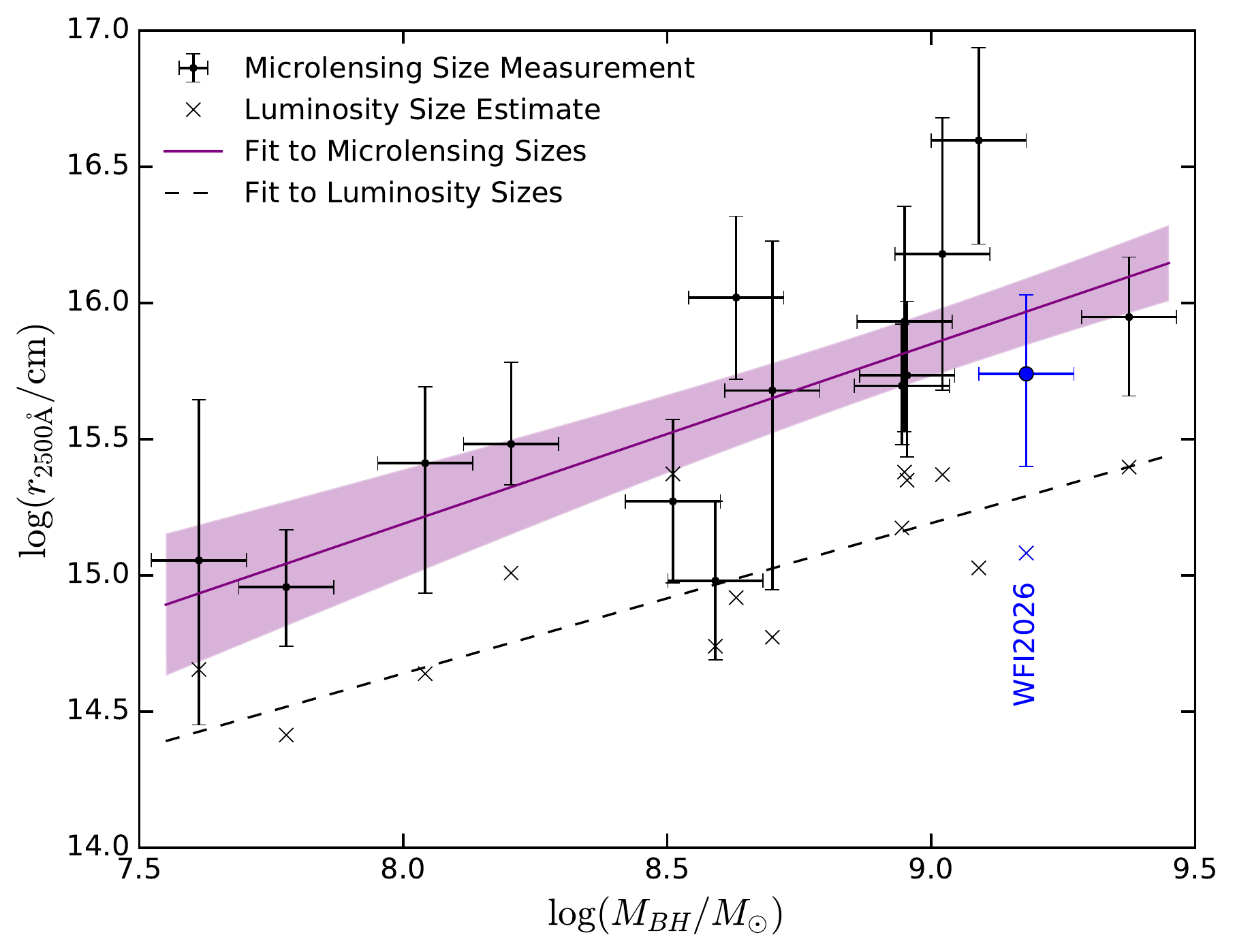}
\caption{Quasar accretion disk sizes scaled to $\lambda_{\text{rest}} = 2500\text{\AA}$ plotted as a function of the central black hole masses.  WFI2026 is highlighted in blue while the other available microlensing size measurements are shown as black dots \citep{koch2004a, morg2008a, dai2010a, morg2010a, hain2012a, morg2012a, hain2013a, macl2015a,morg2018a}.  The best-fit line from \citet{morg2018a} is shown in purple with $1\sigma$ errorbars encompassed by the purple band.  The luminosity-based size estimates are shown with black diagonal crosses with the best fit indicated by the black dotted line.}
\label{fig:mbh_r}
\end{figure}


The key measurement we have presented in this study is the accretion disk size of WFI2026, shown in Figure~\ref{fig:rhat}.  Our findings with and without a prior on microlensing mass are consistent so this result is essentially independent of the unknown median microlensing mass. 
To compare our measurement to those from other studies on WFI2026, we convert this to a half-light radius, under the thin-disk assumption, to give a value of  $\log\{(r_{1/2}/\text{cm}) [\cos(i)/0.5]^{-1/2}\} = 16.13^{+0.34}_{-0.29}$.   Our estimate is smaller, but consistent with the findings of \citet{blac2011a},  $\log (r_{1/2}/\text{cm}) = 16.46 \pm 0.32$, and larger than the estimate $\log (r_{1/2}/\text{cm}) < 16$ found by \citet{bate2018a}, but again, consistent within statistical bounds.  The estimate from \citet{bate2018a} adjusts the half-light radius based on their empirical temperature slope rather than thin-disk slope.  If instead, we adjust their scale size based on thin-disk scaling, their estimate increases to $\log (r_{1/2}/\text{cm}) < 16.14$, fully consistent with our measurement.

Although we found a well-constrained measurement of the physical accretion disk size, we did encounter a bimodal distribution in the effective radius and source velocity which was mitigated by the degeneracy between $\hat{v}_e$ and $\hat{r}_s$.  In any case, the application of a velocity prior validated these results.

We also reported the first spectroscopic measurement of the central black hole mass based on the \ion{C}{4} line width and the relation from \citet{vest2006a}.  This is a relatively large central black hole, the second most massive in our sample of 15 quasars.  Our estimate of $\log(M_{BH}/M_{\odot}) = 9.18^{+0.39}_{-0.34}$ is larger than the bolometric luminosity estimate from \citet{blac2011a} of $\log(M_{BH}/M_{\odot}) = 8.90$ in accordance with their findings that luminosity-based estimates are systematically smaller than virial estimates that also use the broad emission line width.

With our black hole mass estimate, we compared our results from WFI2026 to the $r_{\mu}$ vs. $M_{\text{BH}}$ relation from \citet{morg2010a, morg2018a}.  To match this study, we shifted our scale radius to $\lambda_{\text{rest}}=2500\,\text{\AA}$, assuming a thin disk model, which gave $\log(r_s/\text{cm} [\cos(i)/0.5]^{-1/2}) = 15.86^{+0.34}_{-0.29}$.  This value is fully consistent with the estimate of $\log(r_{2500}/\text{cm}) = 15.97 \pm 0.12$ predicted by the accretion disk size -- $M_{\text{BH}}$ relation of \citet{morg2018a},  as can be seen in Figure~\ref{fig:mbh_r}.

Figure~\ref{fig:mbh_r} also displays that the microlensing sizes are systematically larger than the theoretical thin disk sizes one would predict using standard thin disk theory (see \citet{morg2010a} for details).  Following the same approach and using the Magellan $i$-band flux from \citet{morg2004a} we estimated the luminosity size for WFI2026.  Adopting an inclination angle of $60\degr$, we found $\log\{(r_L/\text{cm})[\cos(i)/0.5]^{-1/2}\} = 15.08 \pm 0.13$, when re-scaled to a rest-frame wavelength of $2500\,\text{\AA}$.  This estimate is smaller than the microlensing size measurements by $0.66\pm0.33\,\text{dex}$, similar to the offset reported in previous microlensing studies.

The general consistency of these fifteen studies offers an opportunity to infer host quasar properties.  In a forthcoming work, we will use the observed size offset and the framework developed in \citet{morg2010a} to provide a robust observational constraint on the quasar accretion disk temperature profile.

\acknowledgments

This material is based upon work supported by the National Science Foundation under grant AST-1614018 to M.A.C. and C.W.M.

COSMOGRAIL is made possible thanks to the continuous work of all observers and technical staff obtaining the monitoring observations, in particular, at the Swiss Leonhard Euler telescope at La Silla Observatory, which is supported by the Swiss National Science Foundation.  M.M., F.C. and V.B. acknowledge support from the Swiss National Science Foundation (SNSF) and through European Research Council (ERC) under the European Union’s Horizon 2020 research and innovation programme (COSMICLENS: grant agreement No 787886). 

MCB gratefully acknowledges support from the National Science Foundation through CAREER grant AST-1253702 to Georgia State University.

We are grateful to Christopher S. Kochanek for the use of his microlensing analysis code.

\facilities{\textsl{HST} (NICMOS),  VLT:Kueyen, Euler:1.2m}

\software{Anaconda, \texttt{LENSMODEL} \citep{keet2001a},  \texttt{PyCS} \citep{tewe2013a}}

\clearpage

\appendix
\section{WFI2026 Lightcurve Table}
\label{sec:lc_table}

\startlongtable
\begin{deluxetable*}{cccc}
\tablecaption{Reduced lightcurve for WFI2026.  The heliocentric julian date (HJD) given is HJD-2,450,000 days.  Images A=A1+A2, B, and C are given in magnitudes relative to the nearby reference stars.\label{tab:lightcurve}}
\tablehead{
\colhead{HJD (days)} & \colhead{A=A1+A2 (mag)} & \colhead{B (mag)} & \colhead{C (mag)}
}
\startdata
$3125.903$ & $2.641\pm 0.006$ & $4.686\pm 0.014$ & $4.730\pm 0.029$ \\ 
$3126.893$ & $2.654\pm 0.002$ & $4.694\pm 0.002$ & $4.664\pm 0.048$ \\ 
$3149.879$ & $2.674\pm 0.010$ & $4.612\pm 0.037$ & $4.739\pm 0.058$ \\ 
$3151.820$ & $2.642\pm 0.004$ & $4.667\pm 0.003$ & $4.772\pm 0.030$ \\ 
$3153.856$ & $2.657\pm 0.008$ & $4.604\pm 0.006$ & $4.740\pm 0.038$ \\ 
$3156.804$ & $2.655\pm 0.008$ & $4.673\pm 0.034$ & $4.723\pm 0.059$ \\ 
$3158.791$ & $2.652\pm 0.005$ & $4.677\pm 0.010$ & $4.817\pm 0.054$ \\ 
$3160.851$ & $2.645\pm 0.001$ & $4.646\pm 0.003$ & $4.866\pm 0.031$ \\ 
$3162.794$ & $2.651\pm 0.001$ & $4.654\pm 0.008$ & $4.722\pm 0.075$ \\ 
$3183.835$ & $2.657\pm 0.004$ & $4.673\pm 0.011$ & $4.786\pm 0.019$ \\ 
$3194.875$ & $2.646\pm 0.004$ & $4.702\pm 0.009$ & $4.865\pm 0.026$ \\ 
$3197.851$ & $2.627\pm 0.010$ & $4.701\pm 0.037$ & $4.889\pm 0.058$ \\ 
$3203.861$ & $2.627\pm 0.003$ & $4.705\pm 0.011$ & $4.871\pm 0.012$ \\ 
$3226.768$ & $2.630\pm 0.010$ & $4.672\pm 0.037$ & $4.765\pm 0.058$ \\ 
$3241.782$ & $2.645\pm 0.005$ & $4.685\pm 0.009$ & $4.732\pm 0.071$ \\ 
$3245.776$ & $2.614\pm 0.003$ & $4.662\pm 0.016$ & $4.952\pm 0.015$ \\ 
$3248.711$ & $2.621\pm 0.003$ & $4.655\pm 0.021$ & $4.819\pm 0.015$ \\ 
$3294.634$ & $2.564\pm 0.010$ & $4.654\pm 0.037$ & $4.852\pm 0.058$ \\ 
$3296.612$ & $2.555\pm 0.005$ & $4.674\pm 0.011$ & $4.878\pm 0.018$ \\ 
$3302.557$ & $2.563\pm 0.001$ & $4.634\pm 0.008$ & $4.807\pm 0.009$ \\ 
$3303.590$ & $2.566\pm 0.003$ & $4.608\pm 0.021$ & $4.804\pm 0.017$ \\ 
$3309.548$ & $2.557\pm 0.010$ & $4.609\pm 0.037$ & $4.810\pm 0.058$ \\ 
$3310.599$ & $2.570\pm 0.005$ & $4.629\pm 0.005$ & $4.752\pm 0.042$ \\ 
$3329.540$ & $2.538\pm 0.010$ & $4.619\pm 0.037$ & $4.740\pm 0.058$ \\ 
$3334.564$ & $2.556\pm 0.010$ & $4.698\pm 0.037$ & $4.602\pm 0.058$ \\ 
$3346.579$ & $2.549\pm 0.010$ & $4.631\pm 0.037$ & $4.641\pm 0.058$ \\ 
$3353.559$ & $2.566\pm 0.010$ & $4.659\pm 0.037$ & $4.519\pm 0.058$ \\ 
$3431.853$ & $2.490\pm 0.018$ & $4.580\pm 0.005$ & $4.514\pm 0.017$ \\ 
$3432.861$ & $2.513\pm 0.011$ & $4.561\pm 0.040$ & $4.379\pm 0.028$ \\ 
$3433.889$ & $2.496\pm 0.005$ & $4.564\pm 0.023$ & $4.502\pm 0.015$ \\ 
$3434.865$ & $2.491\pm 0.005$ & $4.575\pm 0.006$ & $4.477\pm 0.044$ \\ 
$3435.868$ & $2.507\pm 0.010$ & $4.580\pm 0.037$ & $4.384\pm 0.058$ \\ 
$3436.865$ & $2.528\pm 0.003$ & $4.550\pm 0.018$ & $4.421\pm 0.008$ \\ 
$3442.836$ & $2.501\pm 0.010$ & $4.569\pm 0.037$ & $4.527\pm 0.058$ \\ 
$3450.857$ & $2.533\pm 0.009$ & $4.599\pm 0.009$ & $4.563\pm 0.062$ \\ 
$3458.874$ & $2.533\pm 0.000$ & $4.527\pm 0.018$ & $4.579\pm 0.007$ \\ 
$3467.914$ & $2.506\pm 0.010$ & $4.623\pm 0.037$ & $4.595\pm 0.058$ \\ 
$3480.856$ & $2.510\pm 0.005$ & $4.660\pm 0.017$ & $4.676\pm 0.031$ \\ 
$3499.763$ & $2.558\pm 0.010$ & $4.528\pm 0.037$ & $4.761\pm 0.058$ \\ 
$3502.793$ & $2.569\pm 0.005$ & $4.551\pm 0.035$ & $4.694\pm 0.019$ \\ 
$3511.813$ & $2.557\pm 0.006$ & $4.602\pm 0.020$ & $4.727\pm 0.047$ \\ 
$3516.799$ & $2.578\pm 0.010$ & $4.527\pm 0.037$ & $4.716\pm 0.058$ \\ 
$3520.882$ & $2.538\pm 0.010$ & $4.617\pm 0.037$ & $4.852\pm 0.058$ \\ 
$3522.892$ & $2.541\pm 0.000$ & $4.573\pm 0.013$ & $4.813\pm 0.005$ \\ 
$3524.907$ & $2.531\pm 0.001$ & $4.558\pm 0.015$ & $4.734\pm 0.019$ \\ 
$3525.890$ & $2.529\pm 0.003$ & $4.586\pm 0.023$ & $4.802\pm 0.009$ \\ 
$3544.801$ & $2.499\pm 0.010$ & $4.483\pm 0.037$ & $4.863\pm 0.058$ \\ 
$3558.682$ & $2.510\pm 0.007$ & $4.555\pm 0.016$ & $4.613\pm 0.035$ \\ 
$3586.749$ & $2.478\pm 0.005$ & $4.600\pm 0.012$ & $4.871\pm 0.008$ \\ 
$3592.514$ & $2.502\pm 0.011$ & $4.551\pm 0.004$ & $4.653\pm 0.033$ \\ 
$3597.565$ & $2.499\pm 0.010$ & $4.507\pm 0.037$ & $4.668\pm 0.058$ \\ 
$3601.641$ & $2.468\pm 0.010$ & $4.404\pm 0.037$ & $4.820\pm 0.058$ \\ 
$3602.683$ & $2.454\pm 0.001$ & $4.522\pm 0.029$ & $4.761\pm 0.009$ \\ 
$3603.740$ & $2.453\pm 0.003$ & $4.564\pm 0.011$ & $4.749\pm 0.015$ \\ 
$3607.568$ & $2.470\pm 0.007$ & $4.540\pm 0.015$ & $4.667\pm 0.021$ \\ 
$3608.627$ & $2.456\pm 0.002$ & $4.519\pm 0.000$ & $4.690\pm 0.031$ \\ 
$3614.801$ & $2.451\pm 0.010$ & $4.535\pm 0.044$ & $4.529\pm 0.007$ \\ 
$3634.705$ & $2.407\pm 0.010$ & $4.530\pm 0.037$ & $4.534\pm 0.058$ \\ 
$3640.716$ & $2.396\pm 0.005$ & $4.563\pm 0.000$ & $4.518\pm 0.020$ \\ 
$3645.611$ & $2.399\pm 0.004$ & $4.424\pm 0.015$ & $4.525\pm 0.027$ \\ 
$3650.583$ & $2.393\pm 0.003$ & $4.540\pm 0.006$ & $4.645\pm 0.008$ \\ 
$3668.524$ & $2.408\pm 0.001$ & $4.454\pm 0.005$ & $4.570\pm 0.043$ \\ 
$3672.554$ & $2.391\pm 0.003$ & $4.543\pm 0.013$ & $4.609\pm 0.009$ \\ 
$3675.528$ & $2.405\pm 0.002$ & $4.548\pm 0.023$ & $4.443\pm 0.009$ \\ 
$3676.543$ & $2.386\pm 0.010$ & $4.535\pm 0.037$ & $4.583\pm 0.058$ \\ 
$3677.572$ & $2.383\pm 0.003$ & $4.549\pm 0.007$ & $4.589\pm 0.008$ \\ 
$3678.537$ & $2.385\pm 0.001$ & $4.550\pm 0.010$ & $4.629\pm 0.005$ \\ 
$3680.542$ & $2.398\pm 0.003$ & $4.522\pm 0.009$ & $4.500\pm 0.006$ \\ 
$3681.572$ & $2.403\pm 0.010$ & $4.521\pm 0.037$ & $4.459\pm 0.058$ \\ 
$3682.585$ & $2.374\pm 0.005$ & $4.547\pm 0.002$ & $4.544\pm 0.011$ \\ 
$3684.547$ & $2.368\pm 0.003$ & $4.562\pm 0.019$ & $4.564\pm 0.007$ \\ 
$3685.572$ & $2.377\pm 0.010$ & $4.520\pm 0.037$ & $4.551\pm 0.058$ \\ 
$3686.545$ & $2.394\pm 0.000$ & $4.527\pm 0.009$ & $4.505\pm 0.021$ \\ 
$3687.545$ & $2.371\pm 0.001$ & $4.561\pm 0.011$ & $4.579\pm 0.006$ \\ 
$3688.578$ & $2.381\pm 0.010$ & $4.559\pm 0.006$ & $4.589\pm 0.007$ \\ 
$3689.530$ & $2.387\pm 0.001$ & $4.554\pm 0.013$ & $4.468\pm 0.015$ \\ 
$3690.532$ & $2.388\pm 0.005$ & $4.509\pm 0.029$ & $4.476\pm 0.009$ \\ 
$3691.572$ & $2.382\pm 0.002$ & $4.571\pm 0.008$ & $4.484\pm 0.012$ \\ 
$3692.558$ & $2.382\pm 0.002$ & $4.557\pm 0.009$ & $4.528\pm 0.007$ \\ 
$3693.532$ & $2.379\pm 0.010$ & $4.534\pm 0.037$ & $4.594\pm 0.058$ \\ 
$3694.525$ & $2.375\pm 0.004$ & $4.552\pm 0.003$ & $4.584\pm 0.011$ \\ 
$3695.530$ & $2.388\pm 0.010$ & $4.533\pm 0.037$ & $4.509\pm 0.058$ \\ 
$3696.542$ & $2.398\pm 0.002$ & $4.472\pm 0.006$ & $4.564\pm 0.029$ \\ 
$3700.535$ & $2.382\pm 0.006$ & $4.547\pm 0.016$ & $4.537\pm 0.025$ \\ 
$3707.540$ & $2.404\pm 0.003$ & $4.493\pm 0.010$ & $4.489\pm 0.013$ \\ 
$3715.527$ & $2.401\pm 0.009$ & $4.600\pm 0.044$ & $4.501\pm 0.032$ \\ 
$3806.865$ & $2.426\pm 0.010$ & $4.536\pm 0.037$ & $4.511\pm 0.058$ \\ 
$3813.826$ & $2.436\pm 0.005$ & $4.601\pm 0.037$ & $4.342\pm 0.008$ \\ 
$3819.852$ & $2.419\pm 0.003$ & $4.551\pm 0.020$ & $4.523\pm 0.016$ \\ 
$3820.811$ & $2.420\pm 0.010$ & $4.615\pm 0.037$ & $4.416\pm 0.058$ \\ 
$3821.848$ & $2.408\pm 0.003$ & $4.553\pm 0.010$ & $4.641\pm 0.018$ \\ 
$3824.855$ & $2.423\pm 0.003$ & $4.529\pm 0.011$ & $4.491\pm 0.011$ \\ 
$3828.903$ & $2.379\pm 0.001$ & $4.543\pm 0.003$ & $4.589\pm 0.008$ \\ 
$3829.844$ & $2.387\pm 0.010$ & $4.560\pm 0.037$ & $4.522\pm 0.058$ \\ 
$3831.859$ & $2.411\pm 0.009$ & $4.507\pm 0.005$ & $4.445\pm 0.012$ \\ 
$3832.870$ & $2.421\pm 0.002$ & $4.493\pm 0.002$ & $4.450\pm 0.008$ \\ 
$3835.907$ & $2.405\pm 0.002$ & $4.513\pm 0.008$ & $4.546\pm 0.025$ \\ 
$3845.898$ & $2.433\pm 0.004$ & $4.407\pm 0.016$ & $4.560\pm 0.023$ \\ 
$3846.892$ & $2.423\pm 0.003$ & $4.479\pm 0.004$ & $4.508\pm 0.029$ \\ 
$3847.867$ & $2.407\pm 0.002$ & $4.510\pm 0.011$ & $4.579\pm 0.005$ \\ 
$3848.853$ & $2.404\pm 0.002$ & $4.541\pm 0.022$ & $4.494\pm 0.014$ \\ 
$3849.858$ & $2.400\pm 0.003$ & $4.554\pm 0.013$ & $4.572\pm 0.011$ \\ 
$3850.899$ & $2.389\pm 0.001$ & $4.521\pm 0.005$ & $4.629\pm 0.005$ \\ 
$3851.868$ & $2.388\pm 0.001$ & $4.559\pm 0.003$ & $4.640\pm 0.009$ \\ 
$3852.881$ & $2.415\pm 0.004$ & $4.487\pm 0.011$ & $4.559\pm 0.004$ \\ 
$3869.840$ & $2.419\pm 0.004$ & $4.450\pm 0.002$ & $4.576\pm 0.009$ \\ 
$3873.818$ & $2.429\pm 0.001$ & $4.504\pm 0.007$ & $4.538\pm 0.024$ \\ 
$3886.896$ & $2.433\pm 0.003$ & $4.546\pm 0.001$ & $4.685\pm 0.025$ \\ 
$3887.896$ & $2.426\pm 0.002$ & $4.514\pm 0.006$ & $4.704\pm 0.005$ \\ 
$3889.851$ & $2.431\pm 0.005$ & $4.456\pm 0.012$ & $4.670\pm 0.029$ \\ 
$3891.891$ & $2.415\pm 0.002$ & $4.539\pm 0.003$ & $4.692\pm 0.009$ \\ 
$3892.878$ & $2.414\pm 0.005$ & $4.525\pm 0.006$ & $4.715\pm 0.008$ \\ 
$3893.936$ & $2.416\pm 0.001$ & $4.559\pm 0.002$ & $4.606\pm 0.034$ \\ 
$3900.817$ & $2.398\pm 0.007$ & $4.499\pm 0.038$ & $4.613\pm 0.019$ \\ 
$3908.773$ & $2.418\pm 0.001$ & $4.438\pm 0.004$ & $4.666\pm 0.008$ \\ 
$3913.766$ & $2.429\pm 0.002$ & $4.451\pm 0.005$ & $4.673\pm 0.012$ \\ 
$3917.614$ & $2.443\pm 0.010$ & $4.542\pm 0.037$ & $4.526\pm 0.058$ \\ 
$3919.805$ & $2.426\pm 0.010$ & $4.489\pm 0.037$ & $4.709\pm 0.058$ \\ 
$3925.644$ & $2.426\pm 0.002$ & $4.579\pm 0.025$ & $4.507\pm 0.010$ \\ 
$3932.862$ & $2.445\pm 0.008$ & $4.621\pm 0.008$ & $4.630\pm 0.043$ \\ 
$3944.684$ & $2.503\pm 0.003$ & $4.451\pm 0.006$ & $4.670\pm 0.006$ \\ 
$3945.832$ & $2.478\pm 0.002$ & $4.636\pm 0.013$ & $4.781\pm 0.008$ \\ 
$3946.781$ & $2.505\pm 0.003$ & $4.601\pm 0.010$ & $4.708\pm 0.029$ \\ 
$3950.688$ & $2.500\pm 0.000$ & $4.553\pm 0.004$ & $4.805\pm 0.016$ \\ 
$3952.765$ & $2.532\pm 0.010$ & $4.649\pm 0.037$ & $4.742\pm 0.058$ \\ 
$3957.597$ & $2.519\pm 0.010$ & $4.572\pm 0.037$ & $4.712\pm 0.058$ \\ 
$3961.790$ & $2.511\pm 0.003$ & $4.686\pm 0.015$ & $4.812\pm 0.017$ \\ 
$3964.569$ & $2.545\pm 0.001$ & $4.604\pm 0.002$ & $4.528\pm 0.013$ \\ 
$3970.695$ & $2.578\pm 0.004$ & $4.469\pm 0.024$ & $4.520\pm 0.032$ \\ 
$3979.618$ & $2.533\pm 0.004$ & $4.654\pm 0.012$ & $4.828\pm 0.025$ \\ 
$3980.746$ & $2.551\pm 0.007$ & $4.662\pm 0.010$ & $4.664\pm 0.031$ \\ 
$3981.724$ & $2.531\pm 0.001$ & $4.665\pm 0.003$ & $4.899\pm 0.013$ \\ 
$3982.726$ & $2.526\pm 0.004$ & $4.685\pm 0.035$ & $4.844\pm 0.016$ \\ 
$3994.581$ & $2.547\pm 0.010$ & $4.658\pm 0.037$ & $4.829\pm 0.058$ \\ 
$3998.601$ & $2.563\pm 0.008$ & $4.679\pm 0.020$ & $4.693\pm 0.066$ \\ 
$3999.582$ & $2.573\pm 0.010$ & $4.599\pm 0.037$ & $4.773\pm 0.058$ \\ 
$4003.582$ & $2.555\pm 0.001$ & $4.665\pm 0.010$ & $4.855\pm 0.013$ \\ 
$4005.568$ & $2.592\pm 0.000$ & $4.598\pm 0.003$ & $4.704\pm 0.005$ \\ 
$4008.645$ & $2.532\pm 0.004$ & $4.730\pm 0.009$ & $4.920\pm 0.017$ \\ 
$4024.629$ & $2.588\pm 0.002$ & $4.636\pm 0.012$ & $4.637\pm 0.023$ \\ 
$4025.577$ & $2.591\pm 0.004$ & $4.655\pm 0.002$ & $4.581\pm 0.080$ \\ 
$4026.575$ & $2.579\pm 0.010$ & $4.653\pm 0.037$ & $4.679\pm 0.058$ \\ 
$4027.559$ & $2.586\pm 0.006$ & $4.590\pm 0.022$ & $4.650\pm 0.024$ \\ 
$4028.558$ & $2.582\pm 0.007$ & $4.646\pm 0.009$ & $4.589\pm 0.041$ \\ 
$4032.539$ & $2.573\pm 0.001$ & $4.588\pm 0.001$ & $4.667\pm 0.023$ \\ 
$4036.530$ & $2.564\pm 0.001$ & $4.638\pm 0.008$ & $4.825\pm 0.006$ \\ 
$4039.535$ & $2.559\pm 0.010$ & $4.660\pm 0.037$ & $4.742\pm 0.058$ \\ 
$4042.519$ & $2.600\pm 0.000$ & $4.607\pm 0.003$ & $4.546\pm 0.008$ \\ 
$4046.511$ & $2.571\pm 0.008$ & $4.641\pm 0.023$ & $4.753\pm 0.051$ \\ 
$4057.520$ & $2.578\pm 0.004$ & $4.671\pm 0.002$ & $4.653\pm 0.044$ \\ 
$4061.531$ & $2.573\pm 0.003$ & $4.679\pm 0.014$ & $4.750\pm 0.014$ \\ 
$4065.523$ & $2.585\pm 0.007$ & $4.714\pm 0.037$ & $4.607\pm 0.008$ \\ 
$4072.535$ & $2.605\pm 0.003$ & $4.722\pm 0.013$ & $4.628\pm 0.011$ \\ 
$4076.536$ & $2.610\pm 0.017$ & $4.719\pm 0.014$ & $4.622\pm 0.044$ \\ 
$4086.531$ & $2.611\pm 0.010$ & $4.880\pm 0.037$ & $4.626\pm 0.058$ \\ 
$4174.872$ & $2.652\pm 0.005$ & $4.829\pm 0.008$ & $4.652\pm 0.023$ \\ 
$4183.867$ & $2.646\pm 0.006$ & $4.733\pm 0.002$ & $4.682\pm 0.032$ \\ 
$4191.873$ & $2.653\pm 0.003$ & $4.720\pm 0.005$ & $4.707\pm 0.025$ \\ 
$4192.859$ & $2.647\pm 0.002$ & $4.793\pm 0.004$ & $4.729\pm 0.030$ \\ 
$4197.832$ & $2.642\pm 0.005$ & $4.759\pm 0.031$ & $4.543\pm 0.069$ \\ 
$4200.872$ & $2.612\pm 0.002$ & $4.710\pm 0.009$ & $4.688\pm 0.022$ \\ 
$4203.855$ & $2.605\pm 0.001$ & $4.728\pm 0.002$ & $4.681\pm 0.014$ \\ 
$4204.868$ & $2.594\pm 0.010$ & $4.708\pm 0.037$ & $4.763\pm 0.058$ \\ 
$4205.842$ & $2.598\pm 0.002$ & $4.708\pm 0.015$ & $4.734\pm 0.010$ \\ 
$4207.852$ & $2.595\pm 0.007$ & $4.659\pm 0.017$ & $4.846\pm 0.095$ \\ 
$4213.878$ & $2.616\pm 0.002$ & $4.664\pm 0.016$ & $4.648\pm 0.026$ \\ 
$4217.855$ & $2.593\pm 0.002$ & $4.639\pm 0.029$ & $4.762\pm 0.062$ \\ 
$4228.821$ & $2.579\pm 0.010$ & $4.649\pm 0.037$ & $4.606\pm 0.058$ \\ 
$4230.786$ & $2.577\pm 0.007$ & $4.635\pm 0.008$ & $4.762\pm 0.024$ \\ 
$4233.874$ & $2.581\pm 0.002$ & $4.551\pm 0.031$ & $4.751\pm 0.010$ \\ 
$4234.899$ & $2.560\pm 0.008$ & $4.524\pm 0.014$ & $4.849\pm 0.037$ \\ 
$4235.883$ & $2.570\pm 0.001$ & $4.520\pm 0.014$ & $4.746\pm 0.009$ \\ 
$4238.873$ & $2.567\pm 0.003$ & $4.495\pm 0.023$ & $4.760\pm 0.048$ \\ 
$4239.931$ & $2.567\pm 0.010$ & $4.495\pm 0.037$ & $4.686\pm 0.058$ \\ 
$4240.885$ & $2.555\pm 0.003$ & $4.452\pm 0.000$ & $4.692\pm 0.017$ \\ 
$4241.921$ & $2.571\pm 0.001$ & $4.456\pm 0.014$ & $4.623\pm 0.011$ \\ 
$4316.657$ & $2.500\pm 0.003$ & $4.531\pm 0.015$ & $4.760\pm 0.016$ \\ 
$4329.820$ & $2.572\pm 0.004$ & $4.685\pm 0.030$ & $4.471\pm 0.024$ \\ 
$4333.532$ & $2.584\pm 0.009$ & $4.652\pm 0.006$ & $4.573\pm 0.046$ \\ 
$4338.544$ & $2.578\pm 0.003$ & $4.538\pm 0.033$ & $4.898\pm 0.025$ \\ 
$4342.652$ & $2.575\pm 0.005$ & $4.579\pm 0.034$ & $4.789\pm 0.014$ \\ 
$4347.534$ & $2.589\pm 0.006$ & $4.618\pm 0.002$ & $4.677\pm 0.044$ \\ 
$4353.537$ & $2.569\pm 0.003$ & $4.580\pm 0.022$ & $4.800\pm 0.012$ \\ 
$4354.725$ & $2.604\pm 0.010$ & $4.627\pm 0.037$ & $4.490\pm 0.058$ \\ 
$4358.491$ & $2.577\pm 0.009$ & $4.645\pm 0.001$ & $4.623\pm 0.048$ \\ 
$4363.503$ & $2.567\pm 0.003$ & $4.532\pm 0.020$ & $4.707\pm 0.034$ \\ 
$4369.601$ & $2.524\pm 0.001$ & $4.624\pm 0.011$ & $4.804\pm 0.014$ \\ 
$4371.497$ & $2.537\pm 0.010$ & $4.521\pm 0.037$ & $4.739\pm 0.058$ \\ 
$4377.514$ & $2.535\pm 0.002$ & $4.473\pm 0.009$ & $4.735\pm 0.010$ \\ 
$4385.550$ & $2.526\pm 0.000$ & $4.559\pm 0.019$ & $4.728\pm 0.093$ \\ 
$4393.508$ & $2.540\pm 0.002$ & $4.502\pm 0.009$ & $4.696\pm 0.022$ \\ 
$4408.527$ & $2.549\pm 0.005$ & $4.629\pm 0.017$ & $4.649\pm 0.016$ \\ 
$4415.514$ & $2.579\pm 0.007$ & $4.635\pm 0.019$ & $4.544\pm 0.055$ \\ 
$4423.539$ & $2.565\pm 0.008$ & $4.669\pm 0.007$ & $4.590\pm 0.026$ \\ 
$4432.527$ & $2.584\pm 0.003$ & $4.718\pm 0.011$ & $4.590\pm 0.021$ \\ 
$4437.534$ & $2.617\pm 0.014$ & $4.641\pm 0.010$ & $4.420\pm 0.006$ \\ 
$4540.874$ & $2.695\pm 0.006$ & $4.773\pm 0.008$ & $4.744\pm 0.035$ \\ 
$4542.902$ & $2.696\pm 0.010$ & $4.791\pm 0.037$ & $4.674\pm 0.058$ \\ 
$4544.881$ & $2.689\pm 0.007$ & $4.771\pm 0.003$ & $4.725\pm 0.006$ \\ 
$4545.904$ & $2.699\pm 0.013$ & $4.779\pm 0.007$ & $4.624\pm 0.079$ \\ 
$4546.874$ & $2.713\pm 0.007$ & $4.810\pm 0.008$ & $4.594\pm 0.026$ \\ 
$4550.901$ & $2.723\pm 0.004$ & $4.749\pm 0.013$ & $4.612\pm 0.029$ \\ 
$4551.879$ & $2.709\pm 0.010$ & $4.801\pm 0.037$ & $4.634\pm 0.058$ \\ 
$4552.913$ & $2.702\pm 0.010$ & $4.725\pm 0.037$ & $4.709\pm 0.058$ \\ 
$4561.879$ & $2.722\pm 0.005$ & $4.812\pm 0.024$ & $4.740\pm 0.029$ \\ 
$4573.886$ & $2.699\pm 0.010$ & $4.800\pm 0.037$ & $5.049\pm 0.058$ \\ 
$4579.824$ & $2.725\pm 0.010$ & $4.619\pm 0.037$ & $5.072\pm 0.058$ \\ 
$4587.916$ & $2.741\pm 0.010$ & $4.721\pm 0.037$ & $4.829\pm 0.058$ \\ 
$4591.836$ & $2.726\pm 0.000$ & $4.739\pm 0.015$ & $4.899\pm 0.015$ \\ 
$4595.860$ & $2.720\pm 0.006$ & $4.726\pm 0.040$ & $4.815\pm 0.059$ \\ 
$4597.846$ & $2.718\pm 0.003$ & $4.695\pm 0.016$ & $4.929\pm 0.012$ \\ 
$4598.915$ & $2.733\pm 0.010$ & $4.626\pm 0.009$ & $4.859\pm 0.041$ \\ 
$4600.836$ & $2.707\pm 0.010$ & $4.669\pm 0.037$ & $5.020\pm 0.058$ \\ 
$4602.904$ & $2.725\pm 0.010$ & $4.633\pm 0.037$ & $4.945\pm 0.058$ \\ 
$4606.909$ & $2.731\pm 0.001$ & $4.609\pm 0.022$ & $4.814\pm 0.016$ \\ 
$4608.801$ & $2.719\pm 0.001$ & $4.697\pm 0.019$ & $5.013\pm 0.041$ \\ 
$4623.888$ & $2.740\pm 0.001$ & $4.663\pm 0.010$ & $4.800\pm 0.007$ \\ 
$4629.698$ & $2.729\pm 0.001$ & $4.792\pm 0.050$ & $4.724\pm 0.063$ \\ 
$4653.841$ & $2.704\pm 0.003$ & $4.703\pm 0.025$ & $4.870\pm 0.028$ \\ 
$4654.788$ & $2.717\pm 0.004$ & $4.689\pm 0.026$ & $4.734\pm 0.012$ \\ 
$4671.833$ & $2.686\pm 0.006$ & $4.768\pm 0.053$ & $4.639\pm 0.077$ \\ 
$4674.805$ & $2.695\pm 0.000$ & $4.663\pm 0.018$ & $4.695\pm 0.009$ \\ 
$4678.812$ & $2.672\pm 0.004$ & $4.696\pm 0.011$ & $4.735\pm 0.054$ \\ 
$4681.580$ & $2.684\pm 0.004$ & $4.668\pm 0.008$ & $4.739\pm 0.007$ \\ 
$4687.691$ & $2.669\pm 0.002$ & $4.589\pm 0.019$ & $4.794\pm 0.026$ \\ 
$4694.693$ & $2.674\pm 0.005$ & $4.666\pm 0.020$ & $4.825\pm 0.041$ \\ 
$4707.577$ & $2.635\pm 0.008$ & $4.536\pm 0.003$ & $5.047\pm 0.067$ \\ 
$4716.491$ & $2.645\pm 0.010$ & $4.621\pm 0.037$ & $4.840\pm 0.058$ \\ 
$4721.551$ & $2.681\pm 0.008$ & $4.558\pm 0.065$ & $4.572\pm 0.012$ \\ 
$4726.529$ & $2.720\pm 0.008$ & $4.420\pm 0.040$ & $4.491\pm 0.022$ \\ 
$4731.506$ & $2.680\pm 0.002$ & $4.544\pm 0.013$ & $4.670\pm 0.047$ \\ 
$4755.555$ & $2.697\pm 0.002$ & $4.522\pm 0.045$ & $4.812\pm 0.010$ \\ 
$4759.499$ & $2.723\pm 0.018$ & $4.526\pm 0.015$ & $4.704\pm 0.083$ \\ 
$4766.544$ & $2.716\pm 0.002$ & $4.592\pm 0.009$ & $4.942\pm 0.011$ \\ 
$4771.536$ & $2.733\pm 0.010$ & $4.662\pm 0.037$ & $4.806\pm 0.058$ \\ 
$4775.536$ & $2.737\pm 0.004$ & $4.612\pm 0.012$ & $4.849\pm 0.023$ \\ 
$4781.525$ & $2.760\pm 0.003$ & $4.505\pm 0.012$ & $4.850\pm 0.015$ \\ 
$4899.889$ & $2.745\pm 0.003$ & $4.734\pm 0.038$ & $4.787\pm 0.007$ \\ 
$4901.892$ & $2.753\pm 0.010$ & $4.844\pm 0.037$ & $4.662\pm 0.058$ \\ 
$4903.899$ & $2.767\pm 0.005$ & $4.721\pm 0.017$ & $4.593\pm 0.060$ \\ 
$4905.899$ & $2.751\pm 0.003$ & $4.728\pm 0.008$ & $4.725\pm 0.014$ \\ 
$4907.902$ & $2.730\pm 0.000$ & $4.764\pm 0.016$ & $4.770\pm 0.027$ \\ 
$4909.900$ & $2.741\pm 0.008$ & $4.749\pm 0.001$ & $4.612\pm 0.052$ \\ 
$4914.909$ & $2.742\pm 0.004$ & $4.732\pm 0.011$ & $4.659\pm 0.018$ \\ 
$4919.894$ & $2.742\pm 0.011$ & $4.748\pm 0.019$ & $4.588\pm 0.054$ \\ 
$4923.886$ & $2.714\pm 0.010$ & $4.725\pm 0.007$ & $4.649\pm 0.044$ \\ 
$4924.881$ & $2.738\pm 0.008$ & $4.704\pm 0.014$ & $4.544\pm 0.034$ \\ 
$4925.882$ & $2.713\pm 0.003$ & $4.735\pm 0.006$ & $4.648\pm 0.012$ \\ 
$4926.888$ & $2.732\pm 0.000$ & $4.719\pm 0.006$ & $4.549\pm 0.043$ \\ 
$4927.887$ & $2.723\pm 0.008$ & $4.742\pm 0.014$ & $4.543\pm 0.018$ \\ 
$4928.858$ & $2.694\pm 0.013$ & $4.730\pm 0.005$ & $4.701\pm 0.010$ \\ 
$4929.911$ & $2.728\pm 0.007$ & $4.665\pm 0.004$ & $4.564\pm 0.037$ \\ 
$4930.889$ & $2.724\pm 0.011$ & $4.699\pm 0.005$ & $4.574\pm 0.023$ \\ 
$4931.887$ & $2.707\pm 0.006$ & $4.683\pm 0.022$ & $4.697\pm 0.025$ \\ 
$4932.864$ & $2.661\pm 0.003$ & $4.784\pm 0.041$ & $4.929\pm 0.010$ \\ 
$4933.886$ & $2.691\pm 0.002$ & $4.656\pm 0.025$ & $4.756\pm 0.010$ \\ 
$4934.886$ & $2.707\pm 0.003$ & $4.748\pm 0.014$ & $4.644\pm 0.031$ \\ 
$4935.889$ & $2.715\pm 0.003$ & $4.710\pm 0.008$ & $4.563\pm 0.012$ \\ 
$4940.841$ & $2.705\pm 0.010$ & $4.655\pm 0.017$ & $4.635\pm 0.060$ \\ 
$4950.837$ & $2.702\pm 0.010$ & $4.686\pm 0.037$ & $4.649\pm 0.058$ \\ 
$4955.855$ & $2.700\pm 0.000$ & $4.712\pm 0.001$ & $4.716\pm 0.024$ \\ 
$4961.818$ & $2.695\pm 0.001$ & $4.734\pm 0.017$ & $4.724\pm 0.025$ \\ 
$4966.870$ & $2.712\pm 0.001$ & $4.747\pm 0.000$ & $4.764\pm 0.042$ \\ 
$4972.850$ & $2.718\pm 0.003$ & $4.739\pm 0.034$ & $4.787\pm 0.006$ \\ 
$4976.800$ & $2.762\pm 0.002$ & $4.686\pm 0.027$ & $4.788\pm 0.007$ \\ 
$4982.752$ & $2.750\pm 0.006$ & $4.721\pm 0.001$ & $4.789\pm 0.015$ \\ 
$4984.851$ & $2.728\pm 0.010$ & $4.696\pm 0.037$ & $4.926\pm 0.058$ \\ 
$4992.802$ & $2.767\pm 0.009$ & $4.717\pm 0.037$ & $4.718\pm 0.068$ \\ 
$4998.715$ & $2.762\pm 0.010$ & $4.711\pm 0.037$ & $4.801\pm 0.058$ \\ 
$5004.849$ & $2.761\pm 0.010$ & $4.742\pm 0.037$ & $4.737\pm 0.058$ \\ 
$5008.833$ & $2.750\pm 0.005$ & $4.661\pm 0.000$ & $4.901\pm 0.049$ \\ 
$5012.763$ & $2.753\pm 0.002$ & $4.638\pm 0.016$ & $4.847\pm 0.045$ \\ 
$5014.790$ & $2.750\pm 0.001$ & $4.644\pm 0.013$ & $4.842\pm 0.009$ \\ 
$5015.742$ & $2.735\pm 0.002$ & $4.710\pm 0.010$ & $4.892\pm 0.032$ \\ 
$5016.847$ & $2.742\pm 0.006$ & $4.698\pm 0.029$ & $4.889\pm 0.040$ \\ 
$5017.911$ & $2.741\pm 0.002$ & $4.722\pm 0.036$ & $4.846\pm 0.015$ \\ 
$5018.841$ & $2.745\pm 0.010$ & $4.771\pm 0.037$ & $4.744\pm 0.058$ \\ 
$5019.854$ & $2.745\pm 0.005$ & $4.691\pm 0.014$ & $4.925\pm 0.029$ \\ 
$5020.821$ & $2.748\pm 0.003$ & $4.788\pm 0.024$ & $4.800\pm 0.025$ \\ 
$5021.839$ & $2.745\pm 0.005$ & $4.759\pm 0.016$ & $4.801\pm 0.033$ \\ 
$5022.863$ & $2.755\pm 0.001$ & $4.700\pm 0.023$ & $4.811\pm 0.029$ \\ 
$5023.891$ & $2.748\pm 0.006$ & $4.818\pm 0.020$ & $4.740\pm 0.029$ \\ 
$5024.866$ & $2.737\pm 0.003$ & $4.777\pm 0.022$ & $4.776\pm 0.050$ \\ 
$5042.577$ & $2.777\pm 0.006$ & $4.767\pm 0.041$ & $4.663\pm 0.033$ \\ 
$5047.630$ & $2.753\pm 0.007$ & $4.793\pm 0.038$ & $4.850\pm 0.049$ \\ 
$5053.628$ & $2.747\pm 0.009$ & $4.759\pm 0.096$ & $4.752\pm 0.010$ \\ 
$5057.527$ & $2.752\pm 0.005$ & $4.647\pm 0.021$ & $4.746\pm 0.028$ \\ 
$5063.553$ & $2.721\pm 0.000$ & $4.725\pm 0.011$ & $4.803\pm 0.017$ \\ 
$5067.509$ & $2.728\pm 0.008$ & $4.711\pm 0.002$ & $4.718\pm 0.066$ \\ 
$5072.535$ & $2.722\pm 0.010$ & $4.635\pm 0.037$ & $4.745\pm 0.058$ \\ 
$5082.500$ & $2.677\pm 0.010$ & $4.675\pm 0.037$ & $4.746\pm 0.058$ \\ 
$5098.599$ & $2.636\pm 0.000$ & $4.637\pm 0.016$ & $4.811\pm 0.006$ \\ 
$5104.573$ & $2.626\pm 0.008$ & $4.626\pm 0.010$ & $4.702\pm 0.042$ \\ 
$5107.492$ & $2.636\pm 0.003$ & $4.645\pm 0.001$ & $4.648\pm 0.024$ \\ 
$5111.500$ & $2.628\pm 0.002$ & $4.610\pm 0.032$ & $4.800\pm 0.016$ \\ 
$5115.514$ & $2.617\pm 0.010$ & $4.700\pm 0.037$ & $4.801\pm 0.058$ \\ 
$5120.569$ & $2.639\pm 0.011$ & $4.680\pm 0.016$ & $4.722\pm 0.013$ \\ 
$5121.505$ & $2.662\pm 0.004$ & $4.569\pm 0.038$ & $4.763\pm 0.007$ \\ 
$5123.502$ & $2.651\pm 0.011$ & $4.566\pm 0.022$ & $4.806\pm 0.028$ \\ 
$5127.519$ & $2.701\pm 0.010$ & $4.560\pm 0.037$ & $4.654\pm 0.058$ \\ 
$5131.509$ & $2.677\pm 0.010$ & $4.574\pm 0.037$ & $4.731\pm 0.058$ \\ 
$5273.899$ & $2.572\pm 0.002$ & $4.541\pm 0.009$ & $4.634\pm 0.019$ \\ 
$5275.903$ & $2.579\pm 0.002$ & $4.547\pm 0.012$ & $4.522\pm 0.044$ \\ 
$5278.899$ & $2.559\pm 0.005$ & $4.558\pm 0.009$ & $4.636\pm 0.009$ \\ 
$5281.891$ & $2.565\pm 0.002$ & $4.567\pm 0.004$ & $4.641\pm 0.017$ \\ 
$5283.878$ & $2.574\pm 0.001$ & $4.541\pm 0.011$ & $4.597\pm 0.011$ \\ 
$5284.881$ & $2.573\pm 0.003$ & $4.570\pm 0.105$ & $4.610\pm 0.013$ \\ 
$5285.880$ & $2.559\pm 0.001$ & $4.553\pm 0.003$ & $4.639\pm 0.009$ \\ 
$5291.869$ & $2.538\pm 0.004$ & $4.508\pm 0.014$ & $4.567\pm 0.033$ \\ 
$5295.886$ & $2.524\pm 0.002$ & $4.488\pm 0.018$ & $4.601\pm 0.050$ \\ 
$5300.865$ & $2.523\pm 0.010$ & $4.490\pm 0.037$ & $4.527\pm 0.058$ \\ 
$5304.866$ & $2.530\pm 0.002$ & $4.453\pm 0.011$ & $4.518\pm 0.009$ \\ 
$5308.861$ & $2.508\pm 0.005$ & $4.507\pm 0.012$ & $4.557\pm 0.053$ \\ 
$5313.868$ & $2.500\pm 0.009$ & $4.416\pm 0.040$ & $4.556\pm 0.010$ \\ 
$5317.828$ & $2.490\pm 0.005$ & $4.480\pm 0.034$ & $4.479\pm 0.011$ \\ 
$5321.875$ & $2.479\pm 0.003$ & $4.415\pm 0.026$ & $4.615\pm 0.020$ \\ 
$5324.883$ & $2.462\pm 0.003$ & $4.364\pm 0.007$ & $4.701\pm 0.031$ \\ 
$5328.854$ & $2.458\pm 0.010$ & $4.412\pm 0.050$ & $4.674\pm 0.034$ \\ 
$5335.783$ & $2.459\pm 0.002$ & $4.489\pm 0.008$ & $4.490\pm 0.018$ \\ 
$5336.738$ & $2.464\pm 0.008$ & $4.477\pm 0.014$ & $4.469\pm 0.034$ \\ 
$5339.847$ & $2.469\pm 0.004$ & $4.411\pm 0.020$ & $4.515\pm 0.009$ \\ 
$5340.807$ & $2.463\pm 0.004$ & $4.426\pm 0.031$ & $4.547\pm 0.009$ \\ 
$5341.726$ & $2.475\pm 0.000$ & $4.485\pm 0.012$ & $4.431\pm 0.013$ \\ 
$5348.914$ & $2.474\pm 0.010$ & $4.362\pm 0.037$ & $4.548\pm 0.058$ \\ 
$5362.876$ & $2.471\pm 0.001$ & $4.375\pm 0.006$ & $4.570\pm 0.011$ \\ 
$5364.741$ & $2.478\pm 0.000$ & $4.425\pm 0.004$ & $4.558\pm 0.015$ \\ 
$5373.768$ & $2.486\pm 0.003$ & $4.381\pm 0.032$ & $4.644\pm 0.016$ \\ 
$5377.791$ & $2.480\pm 0.006$ & $4.378\pm 0.032$ & $4.541\pm 0.009$ \\ 
$5389.728$ & $2.486\pm 0.004$ & $4.408\pm 0.030$ & $4.643\pm 0.025$ \\ 
$5397.795$ & $2.496\pm 0.003$ & $4.373\pm 0.001$ & $4.551\pm 0.005$ \\ 
$5399.798$ & $2.517\pm 0.006$ & $4.393\pm 0.016$ & $4.451\pm 0.024$ \\ 
$5407.671$ & $2.510\pm 0.010$ & $4.438\pm 0.037$ & $4.487\pm 0.058$ \\ 
$5408.849$ & $2.506\pm 0.003$ & $4.478\pm 0.048$ & $4.530\pm 0.017$ \\ 
$5412.622$ & $2.497\pm 0.001$ & $4.377\pm 0.009$ & $4.624\pm 0.020$ \\ 
$5436.536$ & $2.470\pm 0.007$ & $4.419\pm 0.008$ & $4.521\pm 0.037$ \\ 
$5439.695$ & $2.504\pm 0.003$ & $4.397\pm 0.011$ & $4.429\pm 0.022$ \\ 
$5443.653$ & $2.485\pm 0.001$ & $4.434\pm 0.028$ & $4.528\pm 0.032$ \\ 
$5447.548$ & $2.463\pm 0.002$ & $4.425\pm 0.009$ & $4.609\pm 0.007$ \\ 
$5477.562$ & $2.425\pm 0.006$ & $4.425\pm 0.007$ & $4.544\pm 0.028$ \\ 
$5485.636$ & $2.448\pm 0.002$ & $4.481\pm 0.005$ & $4.549\pm 0.004$ \\ 
$5488.607$ & $2.463\pm 0.002$ & $4.473\pm 0.007$ & $4.484\pm 0.022$ \\ 
$5492.598$ & $2.464\pm 0.006$ & $4.463\pm 0.010$ & $4.566\pm 0.032$ \\ 
$5495.597$ & $2.481\pm 0.002$ & $4.465\pm 0.011$ & $4.504\pm 0.024$ \\ 
$5503.570$ & $2.507\pm 0.003$ & $4.462\pm 0.004$ & $4.524\pm 0.016$ \\ 
$5506.548$ & $2.497\pm 0.005$ & $4.465\pm 0.005$ & $4.571\pm 0.022$ \\ 
$5517.521$ & $2.505\pm 0.009$ & $4.461\pm 0.008$ & $4.469\pm 0.047$ \\ 
$5526.523$ & $2.495\pm 0.003$ & $4.474\pm 0.005$ & $4.535\pm 0.022$ \\ 
$5655.879$ & $2.615\pm 0.007$ & $4.534\pm 0.003$ & $4.672\pm 0.004$ \\ 
$5662.865$ & $2.644\pm 0.006$ & $4.574\pm 0.006$ & $4.605\pm 0.022$ \\ 
$5667.863$ & $2.620\pm 0.008$ & $4.569\pm 0.005$ & $4.711\pm 0.040$ \\ 
$5671.886$ & $2.640\pm 0.004$ & $4.547\pm 0.004$ & $4.643\pm 0.029$ \\ 
$5674.873$ & $2.622\pm 0.002$ & $4.572\pm 0.008$ & $4.729\pm 0.017$ \\ 
$5678.856$ & $2.629\pm 0.005$ & $4.552\pm 0.008$ & $4.704\pm 0.025$ \\ 
$5682.873$ & $2.632\pm 0.003$ & $4.545\pm 0.003$ & $4.641\pm 0.017$ \\ 
$5694.828$ & $2.623\pm 0.005$ & $4.537\pm 0.012$ & $4.615\pm 0.012$ \\ 
$5697.912$ & $2.606\pm 0.004$ & $4.559\pm 0.015$ & $4.841\pm 0.025$ \\ 
$5705.871$ & $2.641\pm 0.007$ & $4.500\pm 0.018$ & $4.697\pm 0.008$ \\ 
$5712.840$ & $2.617\pm 0.001$ & $4.530\pm 0.006$ & $4.684\pm 0.003$ \\ 
$5716.874$ & $2.600\pm 0.004$ & $4.562\pm 0.010$ & $4.748\pm 0.022$ \\ 
$5723.752$ & $2.609\pm 0.003$ & $4.580\pm 0.007$ & $4.704\pm 0.045$ \\ 
$5725.748$ & $2.616\pm 0.002$ & $4.564\pm 0.008$ & $4.640\pm 0.017$ \\ 
$5739.699$ & $2.650\pm 0.016$ & $4.534\pm 0.017$ & $4.621\pm 0.025$ \\ 
$5760.734$ & $2.655\pm 0.002$ & $4.433\pm 0.006$ & $4.617\pm 0.059$ \\ 
$5762.691$ & $2.630\pm 0.008$ & $4.530\pm 0.005$ & $4.719\pm 0.005$ \\ 
$5766.794$ & $2.629\pm 0.002$ & $4.558\pm 0.006$ & $4.729\pm 0.021$ \\ 
$5775.669$ & $2.663\pm 0.012$ & $4.540\pm 0.028$ & $4.547\pm 0.062$ \\ 
$5776.619$ & $2.622\pm 0.001$ & $4.579\pm 0.015$ & $4.713\pm 0.009$ \\ 
$5779.584$ & $2.641\pm 0.006$ & $4.564\pm 0.004$ & $4.626\pm 0.045$ \\ 
$5783.740$ & $2.621\pm 0.003$ & $4.576\pm 0.013$ & $4.789\pm 0.007$ \\ 
$5794.655$ & $2.640\pm 0.002$ & $4.586\pm 0.036$ & $4.773\pm 0.009$ \\ 
$5804.517$ & $2.678\pm 0.007$ & $4.632\pm 0.005$ & $4.646\pm 0.029$ \\ 
$5809.508$ & $2.693\pm 0.005$ & $4.631\pm 0.006$ & $4.651\pm 0.024$ \\ 
$5815.520$ & $2.707\pm 0.008$ & $4.622\pm 0.016$ & $4.758\pm 0.020$ \\ 
$5818.546$ & $2.721\pm 0.008$ & $4.602\pm 0.006$ & $4.723\pm 0.004$ \\ 
$5820.632$ & $2.713\pm 0.004$ & $4.614\pm 0.009$ & $4.833\pm 0.005$ \\ 
$5824.677$ & $2.723\pm 0.004$ & $4.631\pm 0.011$ & $4.767\pm 0.014$ \\ 
$5827.533$ & $2.730\pm 0.004$ & $4.578\pm 0.011$ & $4.658\pm 0.008$ \\ 
$5831.513$ & $2.700\pm 0.004$ & $4.637\pm 0.007$ & $4.781\pm 0.021$ \\ 
$5835.635$ & $2.708\pm 0.007$ & $4.657\pm 0.010$ & $4.788\pm 0.022$ \\ 
$5839.586$ & $2.733\pm 0.006$ & $4.587\pm 0.007$ & $4.741\pm 0.024$ \\ 
$5842.497$ & $2.736\pm 0.002$ & $4.627\pm 0.018$ & $4.759\pm 0.006$ \\ 
$5850.632$ & $2.718\pm 0.007$ & $4.643\pm 0.005$ & $4.829\pm 0.017$ \\ 
$5853.519$ & $2.749\pm 0.003$ & $4.601\pm 0.003$ & $4.730\pm 0.029$ \\ 
$5854.505$ & $2.740\pm 0.002$ & $4.588\pm 0.013$ & $4.763\pm 0.010$ \\ 
$5861.506$ & $2.745\pm 0.002$ & $4.621\pm 0.012$ & $4.831\pm 0.009$ \\ 
$5869.511$ & $2.719\pm 0.003$ & $4.618\pm 0.005$ & $4.779\pm 0.045$ \\ 
$5873.550$ & $2.721\pm 0.008$ & $4.639\pm 0.013$ & $4.706\pm 0.033$ \\ 
$5882.516$ & $2.721\pm 0.006$ & $4.683\pm 0.004$ & $4.694\pm 0.015$ \\ 
$5885.518$ & $2.708\pm 0.005$ & $4.670\pm 0.021$ & $4.756\pm 0.048$ \\ 
$5892.528$ & $2.713\pm 0.013$ & $4.679\pm 0.021$ & $4.853\pm 0.024$ \\ 
$6023.881$ & $2.795\pm 0.002$ & $4.691\pm 0.008$ & $4.710\pm 0.015$ \\ 
$6028.875$ & $2.777\pm 0.004$ & $4.710\pm 0.011$ & $4.852\pm 0.009$ \\ 
$6029.884$ & $2.762\pm 0.004$ & $4.718\pm 0.012$ & $4.918\pm 0.041$ \\ 
$6047.845$ & $2.770\pm 0.005$ & $4.675\pm 0.008$ & $4.702\pm 0.019$ \\ 
$6050.839$ & $2.768\pm 0.005$ & $4.646\pm 0.010$ & $4.753\pm 0.013$ \\ 
$6053.868$ & $2.764\pm 0.004$ & $4.632\pm 0.025$ & $4.730\pm 0.016$ \\ 
$6062.898$ & $2.756\pm 0.002$ & $4.667\pm 0.012$ & $4.956\pm 0.018$ \\ 
$6070.911$ & $2.759\pm 0.004$ & $4.634\pm 0.010$ & $4.931\pm 0.008$ \\ 
$6092.913$ & $2.765\pm 0.005$ & $4.672\pm 0.021$ & $4.867\pm 0.004$ \\ 
$6094.868$ & $2.776\pm 0.001$ & $4.666\pm 0.006$ & $4.863\pm 0.008$ \\ 
$6098.874$ & $2.750\pm 0.006$ & $4.696\pm 0.016$ & $4.961\pm 0.026$ \\ 
$6105.713$ & $2.745\pm 0.001$ & $4.688\pm 0.006$ & $4.862\pm 0.017$ \\ 
$6109.796$ & $2.763\pm 0.007$ & $4.711\pm 0.006$ & $4.875\pm 0.010$ \\ 
$6118.711$ & $2.764\pm 0.001$ & $4.695\pm 0.011$ & $4.893\pm 0.004$ \\ 
$6129.679$ & $2.791\pm 0.005$ & $4.702\pm 0.018$ & $4.860\pm 0.034$ \\ 
$6133.625$ & $2.809\pm 0.002$ & $4.727\pm 0.008$ & $4.823\pm 0.024$ \\ 
$6138.566$ & $2.824\pm 0.010$ & $4.773\pm 0.009$ & $4.761\pm 0.016$ \\ 
$6147.576$ & $2.841\pm 0.009$ & $4.770\pm 0.010$ & $4.886\pm 0.009$ \\ 
$6151.567$ & $2.837\pm 0.012$ & $4.712\pm 0.009$ & $4.960\pm 0.007$ \\ 
$6165.542$ & $2.811\pm 0.003$ & $4.807\pm 0.010$ & $5.096\pm 0.014$ \\ 
$6168.561$ & $2.835\pm 0.006$ & $4.739\pm 0.005$ & $4.982\pm 0.035$ \\ 
$6169.549$ & $2.849\pm 0.002$ & $4.735\pm 0.008$ & $4.899\pm 0.017$ \\ 
$6173.659$ & $2.837\pm 0.003$ & $4.771\pm 0.016$ & $4.989\pm 0.008$ \\ 
$6175.697$ & $2.849\pm 0.002$ & $4.762\pm 0.004$ & $4.948\pm 0.011$ \\ 
$6181.562$ & $2.857\pm 0.005$ & $4.700\pm 0.012$ & $4.914\pm 0.004$ \\ 
$6185.489$ & $2.844\pm 0.007$ & $4.731\pm 0.004$ & $4.905\pm 0.024$ \\ 
$6190.504$ & $2.836\pm 0.005$ & $4.730\pm 0.006$ & $4.960\pm 0.024$ \\ 
$6193.493$ & $2.846\pm 0.006$ & $4.690\pm 0.010$ & $4.824\pm 0.028$ \\ 
$6197.523$ & $2.822\pm 0.002$ & $4.719\pm 0.006$ & $4.973\pm 0.017$ \\ 
$6205.496$ & $2.822\pm 0.011$ & $4.739\pm 0.012$ & $4.914\pm 0.027$ \\ 
$6209.490$ & $2.819\pm 0.002$ & $4.739\pm 0.052$ & $5.004\pm 0.038$ \\ 
$6210.499$ & $2.813\pm 0.007$ & $4.759\pm 0.017$ & $5.012\pm 0.007$ \\ 
$6213.520$ & $2.835\pm 0.013$ & $4.761\pm 0.011$ & $4.814\pm 0.090$ \\ 
$6217.591$ & $2.823\pm 0.002$ & $4.812\pm 0.003$ & $4.978\pm 0.034$ \\ 
$6221.511$ & $2.838\pm 0.010$ & $4.775\pm 0.009$ & $4.953\pm 0.022$ \\ 
$6224.531$ & $2.839\pm 0.002$ & $4.830\pm 0.014$ & $4.948\pm 0.006$ \\ 
$6225.521$ & $2.842\pm 0.006$ & $4.781\pm 0.021$ & $5.048\pm 0.040$ \\ 
$6229.510$ & $2.844\pm 0.012$ & $4.787\pm 0.006$ & $5.042\pm 0.064$ \\ 
$6236.528$ & $2.854\pm 0.002$ & $4.827\pm 0.004$ & $5.072\pm 0.009$ \\ 
$6390.900$ & $2.785\pm 0.004$ & $4.709\pm 0.007$ & $4.795\pm 0.015$ \\ 
$6395.912$ & $2.788\pm 0.004$ & $4.691\pm 0.010$ & $4.819\pm 0.027$ \\ 
$6401.892$ & $2.798\pm 0.007$ & $4.726\pm 0.003$ & $4.840\pm 0.020$ \\ 
$6405.873$ & $2.763\pm 0.006$ & $4.767\pm 0.012$ & $5.028\pm 0.041$ \\ 
$6409.863$ & $2.840\pm 0.008$ & $4.702\pm 0.019$ & $4.713\pm 0.031$ \\ 
$6431.850$ & $2.804\pm 0.004$ & $4.777\pm 0.026$ & $5.002\pm 0.014$ \\ 
$6435.851$ & $2.824\pm 0.002$ & $4.708\pm 0.025$ & $4.846\pm 0.020$ \\ 
$6441.787$ & $2.819\pm 0.004$ & $4.745\pm 0.024$ & $4.898\pm 0.075$ \\ 
$6443.755$ & $2.816\pm 0.004$ & $4.736\pm 0.007$ & $4.746\pm 0.033$ \\ 
$6447.734$ & $2.823\pm 0.008$ & $4.788\pm 0.005$ & $4.789\pm 0.014$ \\ 
$6455.875$ & $2.834\pm 0.002$ & $4.733\pm 0.014$ & $4.949\pm 0.028$ \\ 
$6460.668$ & $2.838\pm 0.005$ & $4.781\pm 0.026$ & $4.871\pm 0.027$ \\ 
$6468.705$ & $2.836\pm 0.009$ & $4.741\pm 0.005$ & $4.896\pm 0.014$ \\ 
$6472.782$ & $2.835\pm 0.009$ & $4.718\pm 0.024$ & $4.940\pm 0.022$ \\ 
$6476.891$ & $2.828\pm 0.007$ & $4.766\pm 0.007$ & $5.002\pm 0.064$ \\ 
$6483.897$ & $2.846\pm 0.011$ & $4.817\pm 0.025$ & $4.889\pm 0.040$ \\ 
$6487.851$ & $2.799\pm 0.001$ & $4.731\pm 0.003$ & $4.874\pm 0.006$ \\ 
$6496.683$ & $2.791\pm 0.006$ & $4.754\pm 0.020$ & $4.890\pm 0.023$ \\ 
$6504.781$ & $2.770\pm 0.006$ & $4.826\pm 0.012$ & $4.948\pm 0.081$ \\ 
$6507.731$ & $2.803\pm 0.001$ & $4.712\pm 0.032$ & $4.875\pm 0.041$ \\ 
$6519.649$ & $2.805\pm 0.004$ & $4.702\pm 0.020$ & $4.936\pm 0.011$ \\ 
$6529.605$ & $2.802\pm 0.004$ & $4.749\pm 0.011$ & $5.036\pm 0.021$ \\ 
$6532.641$ & $2.797\pm 0.003$ & $4.734\pm 0.004$ & $4.933\pm 0.016$ \\ 
$6536.570$ & $2.813\pm 0.001$ & $4.704\pm 0.017$ & $4.913\pm 0.025$ \\ 
$6544.550$ & $2.837\pm 0.002$ & $4.741\pm 0.004$ & $4.847\pm 0.032$ \\ 
$6548.698$ & $2.813\pm 0.003$ & $4.747\pm 0.008$ & $4.923\pm 0.019$ \\ 
$6552.500$ & $2.804\pm 0.002$ & $4.738\pm 0.032$ & $4.970\pm 0.072$ \\ 
$6556.492$ & $2.843\pm 0.008$ & $4.733\pm 0.006$ & $4.785\pm 0.078$ \\ 
$6569.497$ & $2.797\pm 0.003$ & $4.731\pm 0.007$ & $4.834\pm 0.005$ \\ 
$6576.500$ & $2.805\pm 0.001$ & $4.691\pm 0.022$ & $4.907\pm 0.004$ \\ 
$6580.507$ & $2.790\pm 0.005$ & $4.727\pm 0.014$ & $4.948\pm 0.020$ \\ 
$6581.581$ & $2.805\pm 0.002$ & $4.743\pm 0.014$ & $4.876\pm 0.016$ \\ 
$6584.605$ & $2.810\pm 0.002$ & $4.750\pm 0.005$ & $4.890\pm 0.023$ \\ 
$6592.565$ & $2.829\pm 0.003$ & $4.782\pm 0.016$ & $4.911\pm 0.026$ \\ 
$6604.514$ & $2.858\pm 0.003$ & $4.763\pm 0.007$ & $4.849\pm 0.014$ \\ 
$6609.516$ & $2.842\pm 0.004$ & $4.742\pm 0.006$ & $4.995\pm 0.016$ \\ 
$6765.887$ & $3.037\pm 0.011$ & $4.902\pm 0.016$ & $4.907\pm 0.054$ \\ 
$6775.875$ & $3.021\pm 0.002$ & $4.862\pm 0.003$ & $4.888\pm 0.013$ \\ 
$6781.830$ & $2.983\pm 0.004$ & $4.866\pm 0.020$ & $5.009\pm 0.025$ \\ 
$6789.876$ & $2.964\pm 0.001$ & $4.803\pm 0.011$ & $5.095\pm 0.031$ \\ 
$6805.808$ & $2.914\pm 0.002$ & $4.787\pm 0.003$ & $4.995\pm 0.018$ \\ 
$6814.872$ & $2.874\pm 0.001$ & $4.734\pm 0.031$ & $5.049\pm 0.014$ \\ 
$6822.683$ & $2.872\pm 0.004$ & $4.811\pm 0.011$ & $4.890\pm 0.014$ \\ 
$6834.718$ & $2.899\pm 0.005$ & $4.804\pm 0.006$ & $4.817\pm 0.022$ \\ 
$6846.801$ & $2.888\pm 0.005$ & $4.733\pm 0.012$ & $4.986\pm 0.009$ \\ 
$6863.610$ & $2.926\pm 0.007$ & $4.830\pm 0.004$ & $4.730\pm 0.044$ \\ 
$6870.693$ & $2.899\pm 0.005$ & $4.820\pm 0.018$ & $5.022\pm 0.014$ \\ 
$6878.565$ & $2.935\pm 0.002$ & $4.858\pm 0.012$ & $4.831\pm 0.022$ \\ 
$6884.734$ & $2.917\pm 0.004$ & $4.787\pm 0.011$ & $5.007\pm 0.028$ \\ 
$6896.518$ & $2.919\pm 0.002$ & $4.871\pm 0.009$ & $4.943\pm 0.008$ \\ 
$6900.488$ & $2.971\pm 0.004$ & $4.813\pm 0.008$ & $4.791\pm 0.015$ \\ 
$6904.534$ & $2.956\pm 0.002$ & $4.842\pm 0.013$ & $4.872\pm 0.014$ \\ 
$6908.504$ & $2.941\pm 0.005$ & $4.845\pm 0.014$ & $4.930\pm 0.007$ \\ 
$6930.565$ & $2.953\pm 0.002$ & $4.830\pm 0.016$ & $5.060\pm 0.011$ \\ 
$6947.502$ & $2.975\pm 0.004$ & $4.811\pm 0.032$ & $5.046\pm 0.008$ \\ 
$6950.585$ & $2.979\pm 0.001$ & $4.909\pm 0.004$ & $4.994\pm 0.016$ \\ 
$6959.512$ & $2.981\pm 0.008$ & $4.951\pm 0.006$ & $5.116\pm 0.028$ \\ 
$6963.563$ & $2.995\pm 0.012$ & $4.947\pm 0.028$ & $5.005\pm 0.062$ \\ 
$6966.522$ & $2.966\pm 0.012$ & $4.962\pm 0.008$ & $5.238\pm 0.056$ \\ 
$6973.545$ & $2.992\pm 0.010$ & $4.945\pm 0.011$ & $5.095\pm 0.042$ \\ 
$6974.552$ & $2.990\pm 0.003$ & $4.945\pm 0.004$ & $5.143\pm 0.011$ \\ 
$6982.526$ & $3.002\pm 0.012$ & $4.968\pm 0.009$ & $5.110\pm 0.064$ \\ 
$7113.879$ & $3.062\pm 0.006$ & $4.993\pm 0.016$ & $5.105\pm 0.033$ \\ 
$7116.908$ & $3.039\pm 0.002$ & $4.988\pm 0.024$ & $5.139\pm 0.009$ \\ 
$7123.885$ & $3.062\pm 0.003$ & $5.004\pm 0.008$ & $5.025\pm 0.016$ \\ 
$7130.875$ & $3.067\pm 0.011$ & $5.018\pm 0.010$ & $5.011\pm 0.069$ \\ 
$7138.834$ & $3.073\pm 0.008$ & $5.040\pm 0.010$ & $5.122\pm 0.023$ \\ 
$7141.829$ & $3.087\pm 0.003$ & $5.036\pm 0.012$ & $5.003\pm 0.018$ \\ 
$7145.820$ & $3.109\pm 0.007$ & $5.024\pm 0.012$ & $5.004\pm 0.035$ \\ 
$7150.778$ & $3.131\pm 0.005$ & $5.018\pm 0.015$ & $4.957\pm 0.026$ \\ 
$7157.864$ & $3.114\pm 0.006$ & $4.983\pm 0.008$ & $5.130\pm 0.031$ \\ 
$7170.786$ & $3.113\pm 0.002$ & $5.081\pm 0.008$ & $5.244\pm 0.018$ \\ 
$7189.885$ & $3.140\pm 0.002$ & $5.042\pm 0.007$ & $5.218\pm 0.009$ \\ 
$7196.852$ & $3.159\pm 0.006$ & $4.991\pm 0.006$ & $5.203\pm 0.022$ \\ 
$7200.704$ & $3.154\pm 0.005$ & $5.018\pm 0.010$ & $5.151\pm 0.027$ \\ 
$7209.809$ & $3.136\pm 0.010$ & $5.032\pm 0.035$ & $5.301\pm 0.067$ \\ 
$7219.687$ & $3.141\pm 0.012$ & $4.985\pm 0.028$ & $5.127\pm 0.062$ \\ 
$7223.757$ & $3.102\pm 0.001$ & $5.006\pm 0.017$ & $5.320\pm 0.041$ \\ 
$7227.582$ & $3.115\pm 0.004$ & $4.984\pm 0.006$ & $5.127\pm 0.023$ \\ 
$7235.725$ & $3.079\pm 0.005$ & $4.911\pm 0.021$ & $5.271\pm 0.025$ \\ 
$7252.566$ & $3.048\pm 0.001$ & $4.952\pm 0.007$ & $5.099\pm 0.008$ \\ 
$7258.541$ & $3.032\pm 0.002$ & $4.957\pm 0.007$ & $5.103\pm 0.005$ \\ 
$7263.541$ & $3.026\pm 0.013$ & $5.003\pm 0.008$ & $5.130\pm 0.112$ \\ 
$7267.506$ & $3.026\pm 0.002$ & $4.969\pm 0.007$ & $5.141\pm 0.004$ \\ 
$7270.514$ & $3.038\pm 0.002$ & $4.939\pm 0.007$ & $5.079\pm 0.009$ \\ 
$7274.503$ & $3.050\pm 0.010$ & $4.949\pm 0.011$ & $5.003\pm 0.024$ \\ 
$7278.498$ & $3.026\pm 0.001$ & $4.939\pm 0.009$ & $5.141\pm 0.017$ \\ 
$7287.508$ & $3.018\pm 0.010$ & $4.944\pm 0.005$ & $5.219\pm 0.066$ \\ 
$7293.605$ & $3.021\pm 0.001$ & $4.910\pm 0.009$ & $5.146\pm 0.004$ \\ 
$7297.540$ & $3.004\pm 0.002$ & $4.889\pm 0.008$ & $5.220\pm 0.007$ \\ 
$7301.526$ & $2.997\pm 0.009$ & $4.869\pm 0.017$ & $5.179\pm 0.013$ \\ 
$7311.590$ & $2.993\pm 0.002$ & $4.912\pm 0.009$ & $5.107\pm 0.021$ \\ 
$7324.572$ & $2.988\pm 0.006$ & $4.913\pm 0.006$ & $5.040\pm 0.015$ \\ 
$7475.879$ & $2.894\pm 0.008$ & $4.849\pm 0.006$ & $4.924\pm 0.025$ \\ 
$7486.892$ & $2.967\pm 0.023$ & $4.884\pm 0.018$ & $4.950\pm 0.045$ \\ 
$7495.869$ & $2.929\pm 0.008$ & $4.844\pm 0.004$ & $5.015\pm 0.048$ \\ 
$7499.826$ & $2.965\pm 0.013$ & $4.878\pm 0.029$ & $4.900\pm 0.016$ \\ 
$7511.873$ & $2.932\pm 0.005$ & $4.856\pm 0.018$ & $4.879\pm 0.033$ \\ 
$7528.927$ & $2.952\pm 0.003$ & $4.833\pm 0.004$ & $5.015\pm 0.053$ \\ 
$7536.912$ & $2.932\pm 0.003$ & $4.827\pm 0.010$ & $5.062\pm 0.040$ \\ 
$7566.845$ & $2.941\pm 0.002$ & $4.855\pm 0.010$ & $4.947\pm 0.020$ \\ 
$7579.665$ & $2.914\pm 0.018$ & $4.828\pm 0.014$ & $4.897\pm 0.092$ \\ 
$7590.653$ & $2.919\pm 0.003$ & $4.860\pm 0.006$ & $4.849\pm 0.027$ \\ 
$7593.579$ & $2.933\pm 0.008$ & $4.872\pm 0.006$ & $4.839\pm 0.056$ \\ 
$7605.624$ & $2.933\pm 0.002$ & $4.817\pm 0.004$ & $4.928\pm 0.015$ \\ 
$7609.533$ & $2.941\pm 0.003$ & $4.866\pm 0.006$ & $4.875\pm 0.044$ \\ 
$7622.650$ & $2.905\pm 0.002$ & $4.863\pm 0.017$ & $5.100\pm 0.006$ \\ 
$7624.548$ & $2.918\pm 0.004$ & $4.890\pm 0.007$ & $4.963\pm 0.057$ \\ 
$7629.631$ & $2.919\pm 0.004$ & $4.869\pm 0.061$ & $5.076\pm 0.010$ \\ 
$7632.655$ & $2.925\pm 0.004$ & $4.843\pm 0.014$ & $5.047\pm 0.010$ \\ 
$7636.513$ & $2.870\pm 0.011$ & $4.960\pm 0.030$ & $5.253\pm 0.096$ \\ 
$7651.659$ & $2.944\pm 0.004$ & $4.878\pm 0.006$ & $5.073\pm 0.006$ \\ 
$7655.597$ & $2.935\pm 0.001$ & $4.860\pm 0.012$ & $5.065\pm 0.012$ \\ 
$7663.573$ & $2.950\pm 0.003$ & $4.849\pm 0.009$ & $5.077\pm 0.012$ \\ 
$7671.557$ & $2.970\pm 0.007$ & $4.863\pm 0.004$ & $5.051\pm 0.029$ \\ 
$7683.507$ & $2.916\pm 0.010$ & $4.979\pm 0.015$ & $5.346\pm 0.098$ \\ 
$7687.509$ & $2.962\pm 0.001$ & $4.837\pm 0.012$ & $5.126\pm 0.016$ \\ 
$7696.513$ & $2.961\pm 0.005$ & $4.865\pm 0.003$ & $5.078\pm 0.011$ \\ 
$7717.551$ & $2.986\pm 0.005$ & $4.973\pm 0.009$ & $4.934\pm 0.020$ \\ 
\enddata
\end{deluxetable*}

\bibliographystyle{aasjournal}
\bibliography{usna_bibtex_archive}

\end{document}